\documentclass[aip,cha,numerical,12pt]{revtex4-2}
\usepackage{graphicx}				
\usepackage{epsfig}
\usepackage{amsmath,amsthm} 
\usepackage{color}
\usepackage{verbatim}				
\usepackage{ulem}
\begin{document}
\title{Pressure-tuning of electronic structure of CeTe$_{3}$ probed by femtosecond collective mode spectroscopy}

\author{Chandra V. Kotyada}
\affiliation{Institute of Physics, Johannes Gutenberg-University, Staudingerweg 7, 55128 Mainz, Germany}

\author{Priyanka Yogi}\email{pyogi@uni-mainz.de}
\affiliation{Institute of Physics, Johannes Gutenberg-University, Staudingerweg 7, 55128 Mainz, Germany}

\author{Amon P. Lanz}
\affiliation{Institute of Physics, Johannes Gutenberg-University, Staudingerweg 7, 55128 Mainz, Germany}

\author{Rolf Heid}
\affiliation{Institute for Quantum Materials and Technologies, Karlsruhe Institute of Technology, Kaiserstr. 12, 76131 Karlsruhe, Germany}

\author{Jonas Tauch}
\affiliation{Department of Physics, University of Konstanz, 78457 Konstanz, Germany}

\author{Manuel Obergfell}
\affiliation{Institute of Physics, Johannes Gutenberg-University, Staudingerweg 7, 55128 Mainz, Germany}
\affiliation{Department of Physics, University of Konstanz, 78457 Konstanz, Germany}

\author{Hanjo Schaefer}
\affiliation{Department of Physics, University of Konstanz, 78457 Konstanz, Germany}

\author{Paula Giraldo-Gallo}
\affiliation{Department of Applied Physics, Stanford University, Stanford, California 94305, USA}
\affiliation{Department of Physics, Universidad de Los Andes, Bogotá 111711, Colombia}

\author{Ian R. Fisher}
\affiliation{Department of Applied Physics, Stanford University, Stanford, California 94305, USA}
\affiliation{Geballe Laboratory for Advanced Materials, Stanford University, Stanford, California 94305, USA}
\affiliation{Stanford Institute for Materials and Energy Sciences, SLAC National Accelerator Laboratory, 2575 Sand Hill Road, Menlo Park, California 94025, USA}

\author{Alexej Pashkin}
\affiliation{Institute of Ion Beam Physics and Materials Research, Helmholtz-Zentrum Dresden-Rossendorf, 01328 Dresden, Germany}

\author{Jure Demsar}\email{demsar@uni-mainz.de}
\affiliation{Institute of Physics, Johannes Gutenberg-University, Staudingerweg 7, 55128
Mainz, Germany}

\date{August 2026}

\begin{abstract}
We use femtosecond optical spectroscopy to study the evolution of coherent order-parameter dynamics in the prototypical charge-density-wave (CDW) system CeTe$_{3}$ under hydrostatic pressure. The CDW transition temperature decreases from $\sim 570$ K at ambient pressure to near room temperature at $\sim 6$ GPa. The pressure dependence of the order-parameter recovery dynamics indicates enhanced electron–phonon coupling with increasing pressure, implying that CDW suppression is driven predominantly by the reduction in Fermi-surface nesting. Above 7 GPa, no evidence of CDW order is observed down to cryogenic temperatures. Concurrently, the relaxation dynamics exhibit significant slowing down at low-temperatures, consistent with emerging heavy-electron behavior due to pressure-enhanced hybridization between localized Ce 4f levels and itinerant carriers.

\end{abstract}
\maketitle

\section*{Introduction}

While the first experimental observations of Charge Density Waves (CDWs)
occurred in the early 1970s, and CDWs were elaborately investigated already
throughout the 1980s,\cite{gruner1988,Monceau2012} the research in CDW related phenomena gained a renewed interest in the last decade or so. Observations of competition of CDW instabilities with superconductivity in
cuprates,\cite{wise2008,Ghiringhelli2012,torchinsky2013,daSilva2014,miao2021}
nickelates,\cite{tam2022,xu2025} and pnictides\cite{souliou2022,pokharel2022}
interplay of CDW and magnetic orders,\cite{kolincio2017,teng2023,schwark2026}
unconventional CDWs linked to orbital order and loop-currents,
\cite{tan2021,wang2022,cao2023,teng2023,Elmers2025,singh2025} and observations of chiral CDW
orders\cite{yang2022,wang2024} all contributed to enhanced interest in this field of research. Moreover, utilizing femtosecond techniques,
the out-of-equilibrium properties of CDW systems have been investigated,\cite{yusupov2008,schmitt2011} including detailed studies of collective modes,\cite{schaefer2010,warawa2023} photoinduced melting of CDW order\cite{rohwer2011,porer2014,frigge2017} and phase
transitions between different CDW phases,\cite{eichberger2010,haupt2016,danz2021} leading also to observations of
photoinduced metastable hidden phases. \cite{stojchevska2014,vaskivskyi2015,maklar2023}

Any systematic search for new non-equilibrium metastable phases in quantum materials will likely require tuning of competing orders such as charge ordering, magnetism or superconductivity. When considering systems
displaying low temperature structural instabilities, such tuning can be achieved
either by doping or chemical substitution,\cite{sacchetti2006,NLWang2014} strain\cite{tuniz2025,lanz2026} of hydrostatic pressure.\cite{sacchetti2007,xu2025}
Particularly strain and pressure tuning present appealing approaches, since
variation of disorder effects is minimized compared to studies using chemical doping/substitution.

\begin{figure}
[ptbh]
\includegraphics[width=0.95\textwidth]{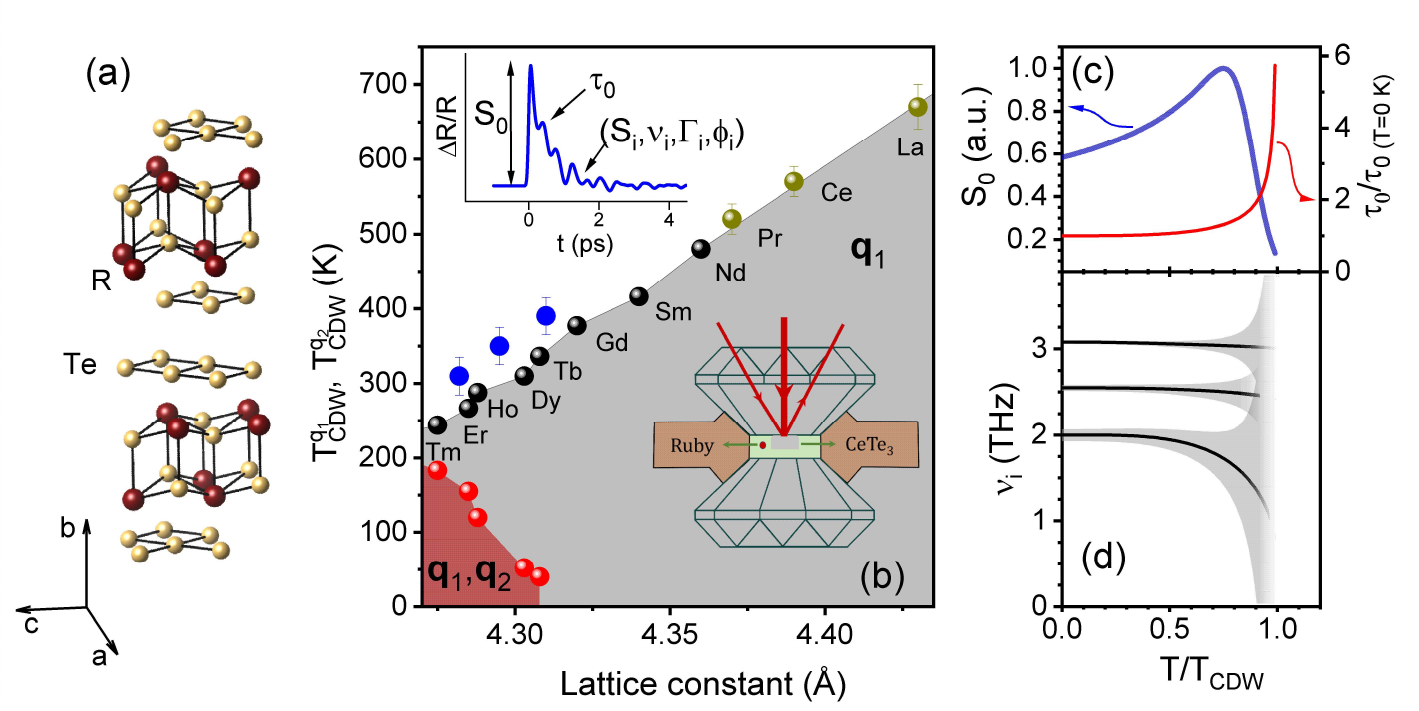}%

\caption{\textbf{Real-time dynamics of order parameter in a CDW system.} (a) Crystal structure of RTe$_{3}$, with red spheres representing the rare-earth elements (R) and the yellow spheres denoting the Te atoms. (b)
Phase diagram of RTe$_{3}$ at ambient pressure, revealing the near linear
dependence of T$_{CDW}^{q_{1}}$ on chemical pressure (adapted
from Ref.\onlinecite{walmsley2020,yumigeta2021}, with T$_{CDW}^{q_{1}}$ for R =
Pr,Ce,La representing extrapolated values. Solid blue circles represent the values of T$_{CDW}^{q_{1}}$ for RTe$_{3}$ extracted from our pressure-dependent data (see Figure \ref{Fig3}). We use the relation between the lattice constant and pressure from Ref. [\onlinecite{sacchetti2009}]. Top inset: Characteristic shape of the reflectivity transient recorded in RTe$_{3}$ systems at temperatures well below T$_{CDW}$
(see, e.g., Ref. [\onlinecite{yusupov2008}]). The response is characterized by an
overdamped mode and several oscillatory modes. (c) Schematic T-dependence of
amplitude (S$_{0}$) and relaxation time ($\tau_{0}$) of the overdamped mode - see Ref. [\onlinecite{schaefer2010}].
(d) Schematic evolution of frequencies (solid lines) and dampings (given by the
shaded region) of the oscillatory modes. These modes become overdamped as
T$_{CDW}$ is approached.}%
\label{Fig1}%
\end{figure}

CeTe$_{3}$ is a member of rare-earth tritelluride series (RTe$_{3}$, R = La--Tm, excluding Eu),
displaying a layered crystal structure made up of double layers of tellurium
(Te) separated by corrugated RTe slabs as shown in Figure \ref{Fig1}(a).
RTe$_{3}$ have orthorhombic symmetry with the long b-axis oriented
perpendicular to the tellurium sheets, and belong to the Cmcm space
group. All members of the series exhibit an unidirectional incommensurate CDW
with a wavevector of q$_{1}$ = (0,0,$\approx2/7 c^{\ast}$) below the
transition temperature T$_{CDW}^{q_{1}}$. For heavier R elements (Tb - Tm), a
second incommensurate CDW with a wavevector q$_{2}$ = ($\approx1/3 a^{\ast}$,0,0)
forms below T$_{CDW}^{q_{2}}$.\cite{walmsley2020,yumigeta2021} As shown in Figure \ref{Fig1}%
(b) T$_{CDW}^{q_{1}}$ varies almost linearly with the lattice constant from
$\approx250$ K (TmTe$_{3}$) to $\approx670$ K (LaTe$_{3}$). The fact that the
formation of CDW in RTe$_{3}$ is primarily driven by Fermi surface nesting and 
momentum-dependent electron-phonon coupling\cite{ARPES2008,eiter2012,maschek2015,maschek2018} makes it highly sensitive to pressure-driven modifications in the electronic structure.\cite{sacchetti2009,li2024} Indeed, several studies of tuning the CDW order
in RTe$_{3}$ by applying hydrostatic pressure have been performed thus
far,\cite{sacchetti2009,Zocco2015,kopaczek2022,yumigeta2022,li2024}
corroborating similarity between the effects of chemical and hydrostatic
pressure on the CDW order, such as pressure-induced reduction of CDW gap,
$\Delta_{CDW}$,\cite{sacchetti2009} and T$_{CDW}^{q_{1}}$.\cite{Zocco2015} Moreover, using hydrostatic pressure, also competing
superconducting phase could be induced in TbTe$_{3}$, GdTe$_{3}$ and
DyTe$_{3}$.\cite{Zocco2015}

Finally, ARPES studies have demonstrated the existence of localized Ce 4f
multiplet just $\approx260$\ meV below the Fermi level in CeTe$_{3}$, which renormalizes the
itinerant bands derived from Te 5p-orbitals,\cite{ARPES2008,Saunot2026} suggesting CeTe$_{3}$ to be a weak
Kondo system.\cite{ru2006,ARPES2008,Saunot2026}

Besides steady-state techniques such as Raman and infrared spectroscopy or ARPES, which act as linear probes of elementary excitations and electronic structure, pump-probe spectroscopy provides a complementary approach for estimating energy gaps associated with electronic orders in solids, as well as structural dynamics through the analysis of carrier relaxation dynamics and coherent lattice dynamics, respectively. The main advantage of pump-probe spectroscopy is its high sensitivity to phase transitions, where energy gaps become small and are therefore difficult to access using linear spectroscopies. Likewise, low-frequency collective modes are readily detected as coherent oscillations in the time domain.

Several real-time studies of coherent order parameter dynamics in RTe$_{3}$ have been
performed to date, ranging from time-resolved optical and ARPES studies in the weak perturbation regime\cite{yusupov2008,schmitt2011,Schmitt2011a,rettig2014} to studies revealing light-induced transformations of CDW
order.\cite{yusupov2010,zong2019,kogar2020,trigo2021,zhou2021} Since our study
focuses on collective response in CeTe$_{3}$, tuned by temperature and the
applied hydrostatic pressure, we first summarize the behavior observed in canonical CDW systems undergoing second order phase transition from a high
temperature metallic phase to an incommensurately modulated CDW ground state,
such as RTe$_{3}$ (R = Ho, Dy, Tb)\cite{yusupov2008} and K$_{0.3}$MoO$_{3}$.
\cite{schaefer2010,warawa2023}

Inset to Figure \ref{Fig1}(b) presents a schematic of a characteristic reflectivity transient
obtained in the CDW phase (in the high temperature metallic state, the response
commonly displays just a rapid sub-ps recovery reflecting a rapid
electron-phonon thermalization). The response is characterized by an
overdamped component with amplitude S$_{0}$ and decay time $\tau_{0}$,
typically in the sub-picosecond range, and several damped oscillatory
components with amplitudes S$_{i}$, frequencies $\nu_{i}$, dampings 
$\Gamma_{i}$, and phases $\phi_{i}$, and can be fit in the time-domain by:

\begin{equation}
\frac{\Delta R}{R}=H\left(  \sigma,t\right)  \left[  S_{0}\cdot e^{-t/\tau
_{0}} + B + \sum_{i} S_{i}\sin\left(  2\pi\nu_{i}t+\phi_{i}\right)  \exp\left(  -\Gamma_{i}t\right)
\right]  , \label{Eq1}%
\end{equation}

where H($\sigma$,t) is a step function with the rise time $\sigma$ \cite{eichberger2010}($\sigma$
is longer than the experimental time resolution and accounts also for
processes immediately after photoexcitation related to the relaxation of hot
carriers and the reduction of electronic order) and $B$ accounts for slowly decaying bolometric response.

 Figure \ref{Fig1}(c) sketches the T-dependence of the fast overdamped
response, with S$_{0}$ vanishing as T$_{CDW}$ is approached, concomitant with
a divergence in $\tau_{0}$. This overdamped component has been attributed to a
recovery of the electronic part of the order parameter, $\Delta$, with
$\tau_{0}^{-1}(T\rightarrow T_{CDW})\propto\Delta^{2}(T)$ and $\tau_{0}%
^{-1}(T=0)$ being proportional to the strength of the electron-phonon coupling.\cite{schaefer2010,schaefer2014,warawa2023} Figure \ref{Fig1}(d) presents the T-dependence of oscillatory modes, which show softening and an increased damping as $T\rightarrow T_{CDW}$. These modes,
appearing only in the low temperature CDW phase, can also be observed in spontaneous
Raman scattering,\cite{sagar2008,yumigeta2022} albeit their spectral weights are usually small compared to the regular Raman active phonons of the parent high temperature phase. These CDW-induced $q\approx0$ Raman active
modes can be naturally identified as CDW amplitude modes, originating from
linear coupling of the electronic order to q$_{CDW}-$phonons of the high
temperature phase.\cite{schaefer2010,schaefer2014,warawa2023,hansen2023,Rice1978} Similarly, new infrared active modes are observed in the CDW state, both in
the time-domain and spectrally resolved studies.
\cite{warawa2023,thomson2017} These infrared active counterparts of Raman
active amplitude modes are identified as phase
modes,\cite{schaefer2010,warawa2023,Rice1978} sometimes referred also as
phase-phonons.\cite{Rice1978}

The damping constants of different amplitude modes and their degree of softening upon approaching $T_{CDW}$ are governed by the individual
coupling strengths of different q$_{CDW}-$phonons to the electronic order and
their overall coupling strength, respectively, the latter also governing the relaxation rate of the electronic part of the order parameter.\cite{schaefer2014,warawa2023} Since all these modes are intrinsically coupled via electronic order, the degree of their softening upon approaching $T_{CDW}$ will also depend on their relative frequencies $\nu_{i}$. E.g., simulations with time-dependent Ginzburg-Landau model have shown that in the case of two modes with 
similar $\nu_{i}$ and comparable coupling strengths to the electronic order,
the lower frequency mode shows a more pronounced softening as the higher
frequency one.\cite{schaefer2014,warawa2023}

Here, we apply time-resolved optical spectroscopy to study the temperature and
pressure dependence of coherent order-parameter dynamics in CeTe$_{3}$, aiming to elucidate the nature of pressure-driven modification of CDW ground state in this canonical CDW system. For hydrostatic pressures below $\sim 6$ GPa, the response is characterized by several amplitude modes, attributed to be a result of linear coupling of the electronic order with normal state phonons at the CDW wavevector. Comparing the frequencies of these modes to those measured by spontaneous Raman scattering (which probe both normal phonons and CDW-induced modes) reveals the selectivity of the pump-probe approach to the amplitude modes. Recording temperature dependences of collective modes at different hydrostatic pressures reveals a pressure-driven reduction of CDW transition temperature from $\sim 570$ K at ambient pressure to near room temperature at $\sim 6$ GPa. Surprisingly, the damping constant of the overdamped electronic mode, which is proportional to the overall coupling strength between the electronic and lattice degrees of freedom of the CDW, reveals an enhancement of the electron-phonon coupling strength with increasing pressure despite the reduction in $\Delta_{CDW\text{,}}$\cite{sacchetti2009} and T$_{CDW}^{q_{1}}$.\cite{Zocco2015} This suggests that the pressure-driven suppression of the CDW order is governed by the pressure-driven changes in the Fermi-surface topology, i.e., reduced Fermi-surface nesting. Increasing the pressure beyond $\sim 6$ GPa results in a collapse of CDW order with no evidence of CDW order down to cryogenic temperatures. Moreover, in this pressure range the electronic relaxation dynamics exhibit a pronounced slowing down upon lowering temperature, which cannot be attributed to a simple metallic behavior. Instead, such a dependence is commonly observed in heavy electron systems where interaction between localized f-moments and conduction electrons leads to hybridization gaps and resulting relaxation bottlenecks,\cite{demsar2006a,demsar2006b} suggesting pressure-driven heavy-electron behavior in CeTe$_3$.

\section*{Results}

We study photoinduced near-infrared (NIR) reflectivity dynamics in CeTe$_{3}$ single crystals as a
function of temperature for several values of pressure up to 7.4 GPa. The
excitation fluence $\approx100$ $\mu$J/cm$^{2}$ (60 fs NIR pulses at
800 nm central frequency) was chosen such that the response remains in the
linear regime, yet enabling a high dynamic range. For pressures below
$\approx6$ GPa the data reveal clear spectroscopic signatures of the CDW
ground state, as sketched in Figure \ref{Fig1}. At 7.4 GPa, however, no
signatures of the CDW are observed down to the lowest temperatures, as shown in Figure S4 in the Supplementary information (SI).

\begin{figure}
[ptb]
\includegraphics[width=0.95\textwidth]{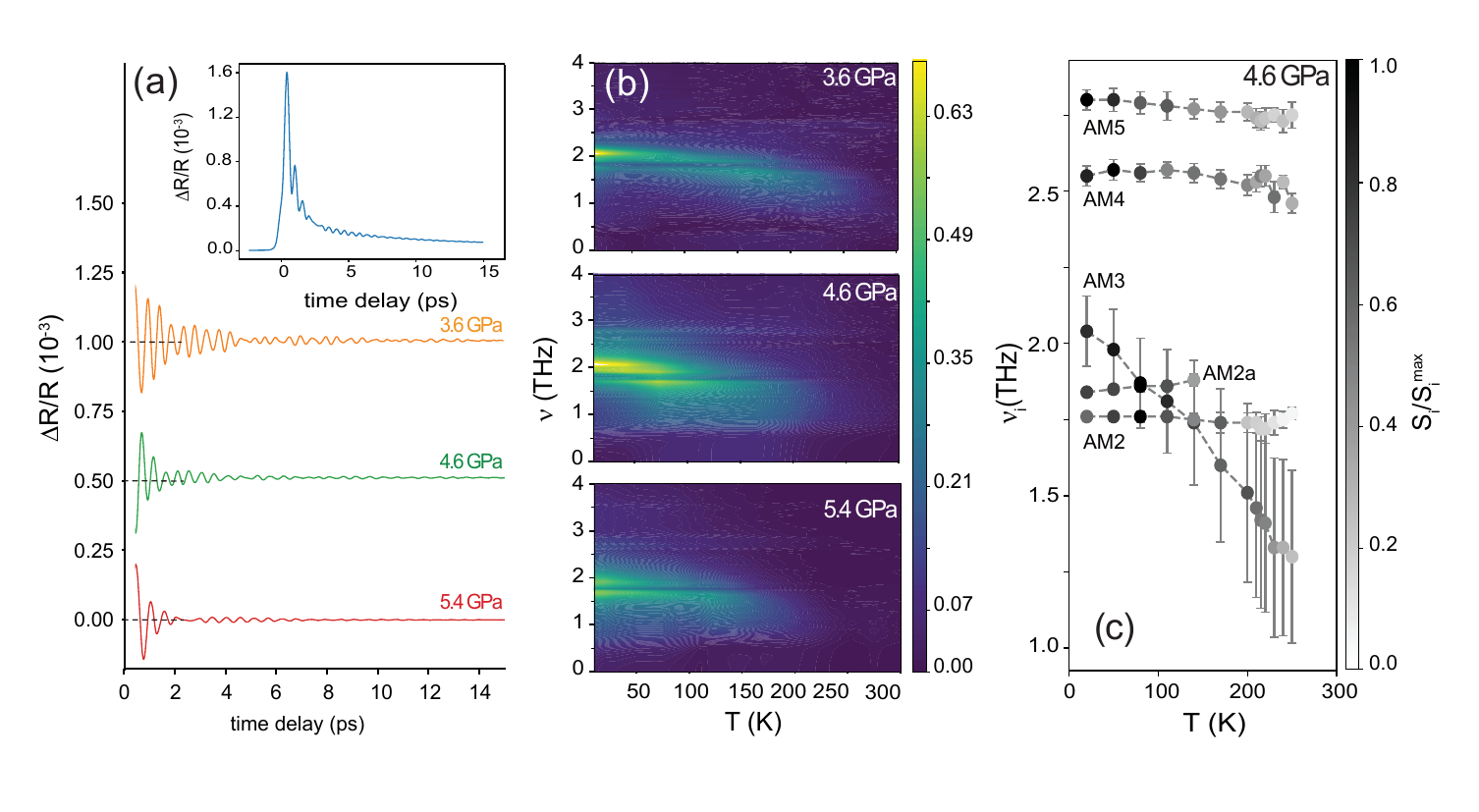}

\caption{\textbf{Temperature dependence of oscillatory collective modes in CeTe$_{3}$
at several pressures}. (a) Oscillatory components of photoinduced NIR
reflectivity traces recorded at 10 K at different pressures, offset for
clarity (inset: NIR reflectivity trace recorded at 10 K at ambient
conditions). (b) T-dependence of the FFT intensity spectrum of the oscillatory
response recorded at 3.6, 4.6 and 5.4 GPa. (c) T-dependence of collective amplitude
modes (AM2-AM5, see also Figure \ref{Fig5}(b)) at 4.6 GPa extracted by real-time fitting of oscillatory traces. Here,
solid symbols denote the mode frequencies, $\nu_{i}$, while error-bars
represent extracted dampings, $\Gamma_{i}$. The mode amplitudes, S$_{i}$,
normalized to the maximum amplitude of the $i$-th mode, S$_{i}^{\max}$, are given
by the shades of gray - see also Figure S1 of the SI.}%
\label{Fig2}%
\end{figure}

We first focus on the data at pressures below 6 GPa. Inset to Figure
\ref{Fig2}(a) presents a reflectivity transient recorded at ambient pressure
at 10 K. The response consists of overdamped and oscillatory components, that is well fit by Eq.\ref{Eq1}. 
Subtracting the overdamped components ($S_{0}\cdot e^{-t/\tau_{0}} + B$) from the raw data singles out the oscillatory response, with traces recorded at 10 K at different pressures shown in Figure \ref{Fig2}(a).

Figure \ref{Fig2}(b) presents the T-dependence of the fast-Fourier transform
(FFT) of the oscillatory response recorded at p = 3.6, 4.6, and 5.4 GPa.
Several oscillatory modes are clearly resolved, displaying softening and
reduction of their oscillator strength upon increasing temperature. The data
recorded at ambient pressure (see Figure S2 in SI) show weak 
temperature dependence in the accessible temperature range (typically 10 - 300
K), which is expected given that T$_{CDW}\approx570$ K at ambient pressure.
The temperature range where oscillatory modes are resolved is clearly shrinking with increasing pressure as seen in Figure \ref{Fig2}(b), suggesting a pressure-driven reduction of T$_{CDW}$.%

To obtain a detailed view of the temperature dependence of modes' oscillator
strengths (S$_{i}$), frequencies ($\nu_{i}$), dampings ($\Gamma_{i}$), and
phases ($\phi_{i}$) we used a real-time fit of the oscillatory response with
the sum of damped oscillators - see \ref{Eq1}. Figure \ref{Fig2}(c) presents the T-dependence of modes' parameters for p = 4.6 GPa. Here the error bars denote the mode dampings $\Gamma_{i}$ (similar as
in Figure \ref{Fig1}(c)) while amplitudes (S$_{i}$), each normalized to its maximum value S$_{i}^{\max}$, are given by the symbol color. It
follows that all modes fade away in a similar fashion upon increasing
temperature, suggesting these are all amplitude modes (AMs) of the
incommensurate CDW in CeTe$_{3}$.

The mode that displays the most prominent softening (AM3 with $\nu_{i}(T\rightarrow
0)\approx2.1$ THz), in the literature usually referred to as the amplitude
mode in RTe$_{3}$, crosses the two lower-lying modes (AM1, AM2) without pronounced
hybridization. The real-time data analysis reveals a smooth evolution of
parameters (S$_{i}$, $\nu_{i}$, $\Gamma_{i}$, $\phi_{i}$) of all modes through
the mode crossings, without any notable transfer of spectral width between the
modes, which is at odds with the reported mode-mixing in RTe$_{3}$ (R = Ho, Dy, Tb).\cite{yusupov2008,yumigeta2021} The lack of significant hybridization
suggests that the underlying displacements of the two lowest-frequency AMs are
nearly orthogonal to the displacements of the dominant AM3 mode.

\begin{figure}
[ptb]
\includegraphics[width=0.6\columnwidth]{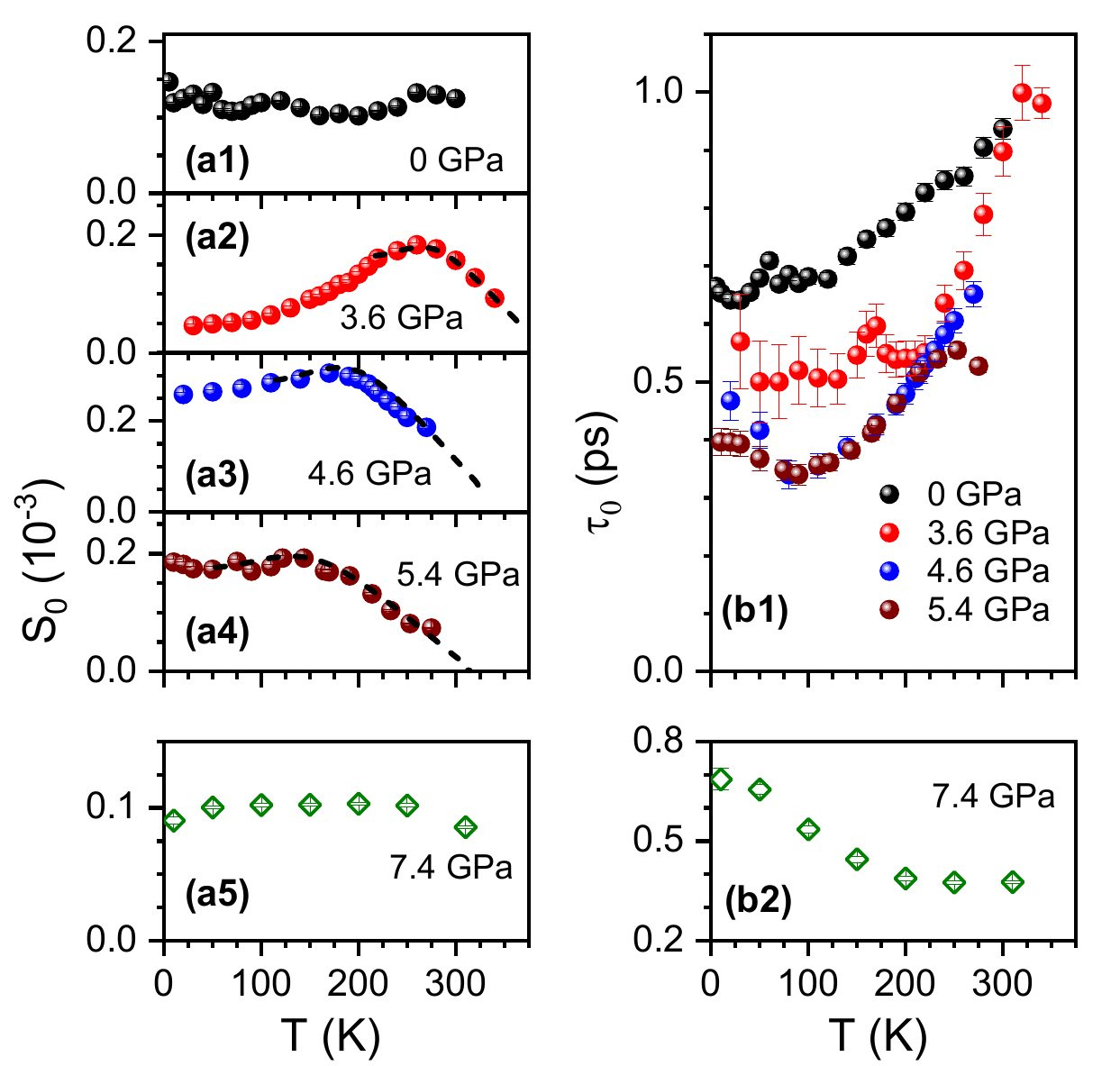}

\caption{\textbf{Temperature dependence of the overdamped response for several pressures.} (a1-a4) amplitude S$_{0}(T)$ and (b1,b2) relaxation time
$\tau_{0}$ recorded at different pressures. At ambient pressure both S$_{0}$
(a1) and $\tau_{0}$ (b1) are nearly constant from 10 to 300 K, which is
expected, given the T$_{c}\approx550$ K (at 0 GPa). At high pressures we
observe the expected reduction in S$_{0}$ (dashed black lines are guides to
the eye), indicating pressure tuning of T$_{c}$, where for p = 5.4 GPa the
T$_{c}$ approaches room temperature. The relaxation time, however, displays an
upturn at temperatures below $\approx150-200$ K, which is not observed at 0
GPa. In terms of the divergence in $\tau_{0}$ upon approaching T$_{c}$, as
observed in RTe$_{3}$ at ambient pressures,\cite{yusupov2008} no clear
signature is observed. Moreover, at room temperature, the relaxation rate
systematically increase with increasing pressure, at odds with the expected
relation $\tau^{-1}\propto\Delta_{CDW}$. Interestingly, at 7.4 GPa (see Figure S4 of the SI) we
observe no spectroscopic evidence of CDW down to cryogenic temperatures.}%
\label{Fig3}%
\end{figure}

While the evolution of AMs with temperature might suggest that for p = 5.4 GPa
$T_{CDW}$ is reduced to well-below room temperature - see Figure \ref{Fig2}(b),
we note that AMs often become overdamped well below $T_{CDW}$. Figure
\ref{Fig3} presents the T-dependence of the overdamped mode recorded at
different pressures. Figure \ref{Fig3}(a1-a5) present S$_{0}(T)$ at different
pressures. At ambient pressure S$_{0}$ shows a weak T-dependence in the
measured T-range, consistent with $T_{CDW}\gg300$ K. Upon increasing pressure
S$_{0}(T)$ do display a characteristic suppression at high temperatures, with
the projected phase transition temperatures (see guides to the eye in Figure
\ref{Fig3}(a2-a4)) suggesting that $T_{CDW}$ is approaching room temperature
at p = 5.4 GPa. This behavior is consistent with earlier studies of
CDW-suppression in pressurized CeTe$_{3}$.\cite{li2024,kopaczek2022,yumigeta2022,Zocco2015,sacchetti2009} 

The relaxation time $\tau_{0}(T)$, shown in Figure \ref{Fig3}(b1), supports the
conclusions based on S$_{0}(T,p)$. For pressures up to 5.4 GPa $\tau_{0}$ is
nearly constant at low temperatures, followed by a gradual increase with
increasing temperature. This behavior is consistent with previous reports in
RTe$_{3}$ \cite{yusupov2008,chen2014,tsuchiya2015} and other CDW systems.\cite{schaefer2010,pokharel2022,demsar2002,Li2023} The fact that no clear
divergence followed by a prompt reduction of relaxation time at $T_{CDW}$ is
observed, such as in (Ho, Dy, Tb)Te$_{3}$,\cite{yusupov2008} underscores the
conclusion based on the S$_{0}(T,p)$ data that $T_{CDW}\gtrsim300$ K even at
5.4 GPa.

Using the relation between the lattice parameter and applied pressure for CeTe$_{3}$,\cite{sacchetti2009} we can compare the results of CDW-tuning with hydrostatic and chemical pressure.\cite{walmsley2020,yumigeta2021} Plotting T$_{CDW}^{q_{1}}(p)$, estimated from the T-dependent studies at 3.6, 4.6 and 5.4 GPa in Figure \ref{Fig1}(b) (solid blue circles), demonstrates similar linear dependence of T$_{CDW}^{q_{1}}$ on the lattice parameter, yet the T$_{CDW}^{q_{1}}$ values obtained under hydrostatic pressure are systematically higher than those obtained by rare-earth substitution. This observation is consistent with the trend observed in optical conductivity studies,\cite{sacchetti2007,sacchetti2009} suggesting optical gaps in pressurized CeTe$_{3}$ systematically higher than those using rare-earth substitution.
Furthermore, no signature of the second incommensurate CDW order with q$_{2}$ is observed in CeTe$_{3}$ under pressure, further pointing out the difference between pressure-tuning and chemical substitution.


At 7.4 GPa the shape of the reflectivity transients is changed (at large time
delays $\frac{\Delta R}{R}<0$, see Figure S4 of the SI), with no oscillatory
modes observed down to 5 K, implying the absence of the CDW ground state. Figure \ref{Fig3}(a5) demonstrates that the amplitude of the fast sub-picosecond
response shows negligible temperature dependence, suggesting a metallic
behavior down to the lowest temperatures. The relaxation time $\tau_{0}$,
shown in Figure \ref{Fig3}(b2), however displays a T-dependence that is at odds
with the temperature dependence of electron-phonon relaxation in metals.\cite{groeneveld1995,Ahn2004} While in conventional metals, the relaxation
time monotonically decreases with decreasing temperature,\cite{groeneveld1995,Ahn2004} the opposite trend is observed in CeTe$_{3}$ at
7.4 GPa. Here, $\tau_{0}$ is nearly constant down to 200 K followed by a
two-fold increase upon lowering the temperature to 10 K. We speculate that the
anomalous low-temperature dependence of $\tau_{0}$ in CeTe$_{3}$ at 7.4 GPa
may be a result of enhanced hybridization between the itinerant carriers
(derived from Te p-orbitals) with localized Ce 4f multiplet under the applied
pressure. Indeed, the localized Ce 4f multiplet just $\approx260$\ meV below
the Fermi level was found to renormalize the itinerant bands already at
ambient pressure.\cite{ARPES2008,Saunot2026} Thus, applying hydrostatic
pressure may enhance hybridization, inducing small hybridization gaps near the
Fermi level. Such small electronic gaps give rise to a carrier
relaxation-bottleneck resulting in slowing down or relaxation at low
temperatures, as observed in many heavy fermion systems.\cite{demsar2006a,demsar2006b,pokharel2021} Indeed, recent scanning tunnelling
spectroscopy on $\beta$-UTe$_{3}$ with spatially-extended 5f orbitals shows
characteristic features of the heavy fermion state and the disappearance of
Fermi-surface nesting.\cite{tuvia2026}


S$_{0}(T,p)$ presented in Figure \ref{Fig3} implies a continuous decrease of
T$_{CDW}$ up to 5.4 GPa, consistent with a continuous suppression of the CDW
gap with pressure.\cite{sacchetti2007} However, it follows from Figure
\ref{Fig3}(b1) that the pressure driven gap reduction results in a decrease of
relaxation time $\tau_{0}$ instead of the expected increase. To address this
aspect, we performed pressure dependent transient reflectivity measurements at room
temperature using finer pressure steps.%

\begin{figure}
[b]

\includegraphics[width=0.6\columnwidth]{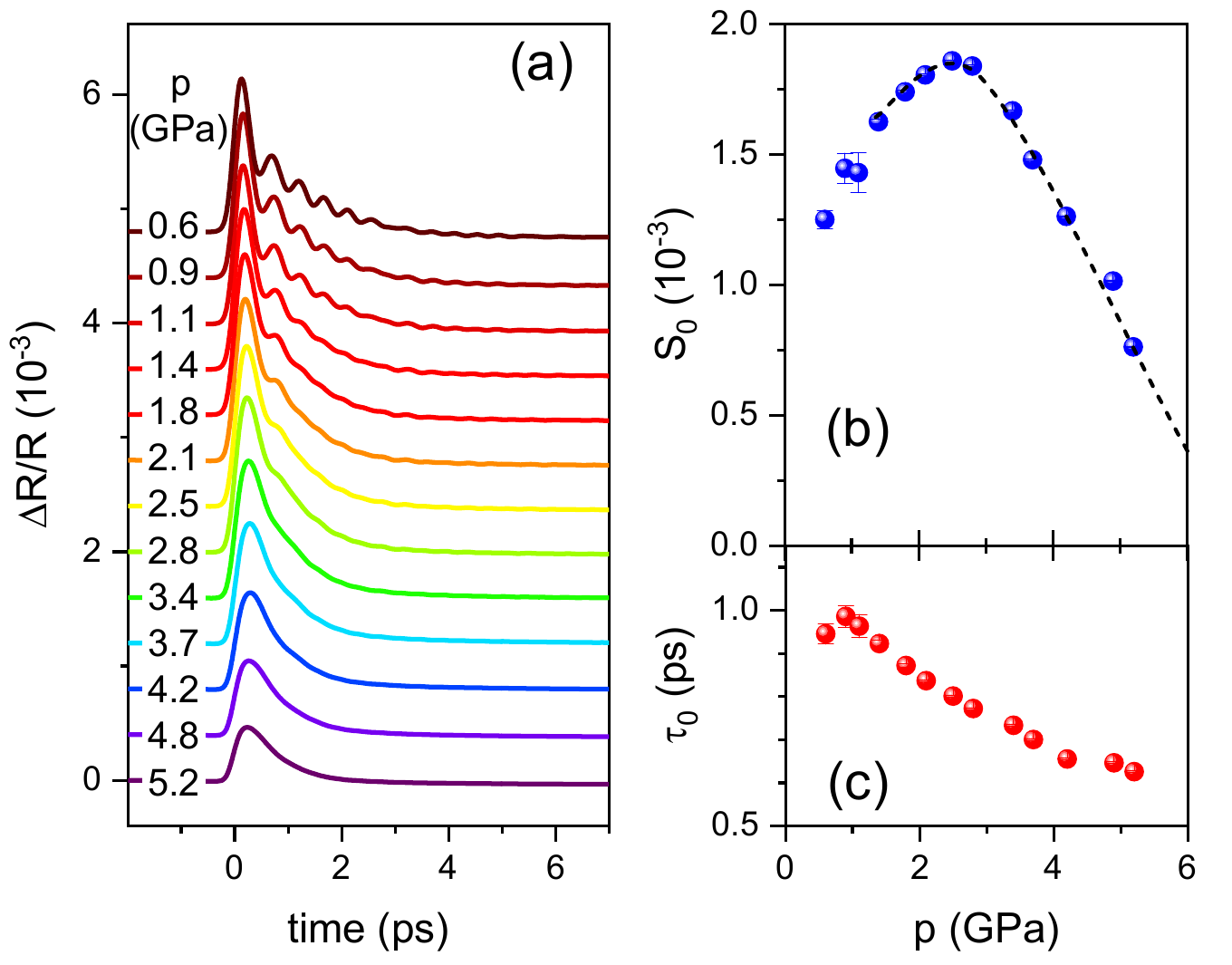}

\caption{\textbf{Detailed pressure dependence of collective response at room temperature.} (a) Reflectivity transients as a function of pressure recorded at
room temperature. (b) The strength of the overdamped signal S$_{0}$ starts
decreasing above $\approx3$ GPa, hinting at the pressure driven collapse of
the CDW at pressures above 6 GPa (dashed line is a guide to the eye). (c) The
recovery time $\tau_{0}$ shows a continuous decrease with increasing pressure,
as already observed in Figure 3(b1).}%
\label{Fig4}%
\end{figure}

Figure \ref{Fig4}(a) presents NIR reflectivity transients recorded at room
temperature as a function of applied pressure, with the extracted S$_{0}(p)$
and $\tau_{0}(p)$ displayed in Figure \ref{Fig4}(b) and (c), respectively.
S$_{0}(p)$ displays the dependence expected for the case where T$_{CDW}$ and
$\Delta_{CDW}$ decrease with increasing pressure. Extrapolating S$_{0}(p)$
beyond 5.2 GPa (dashed line in Figure \ref{Fig4}(b)) suggests CDW order to be
suppressed above $\approx6$ GPa.

Noteworthy, as shown in Figure \ref{Fig4}(c), $\tau_{0}$ decreases as pressure
increases, corroborating the behavior seen in Figure \ref{Fig3}(b1). As
$\tau_{0}^{-1}$ is proportional to the overall electron-phonon coupling strength, this
result appears counter-intuitive. In a Peierls CDW, where $\Delta_{CDW}$
depends on the strength of the electron-phonon coupling, a reduction of
$\Delta_{CDW}$ should result in an increase of $\tau_{0}$ reflecting a
suppression of the electron-phonon coupling. The observed increase in the
electron-phonon coupling strength accompanying CDW suppression thus suggests
that the main effect of pressure is to modify the nesting condition.%

\begin{figure}
[ptb]

\includegraphics[width=0.93\textwidth]{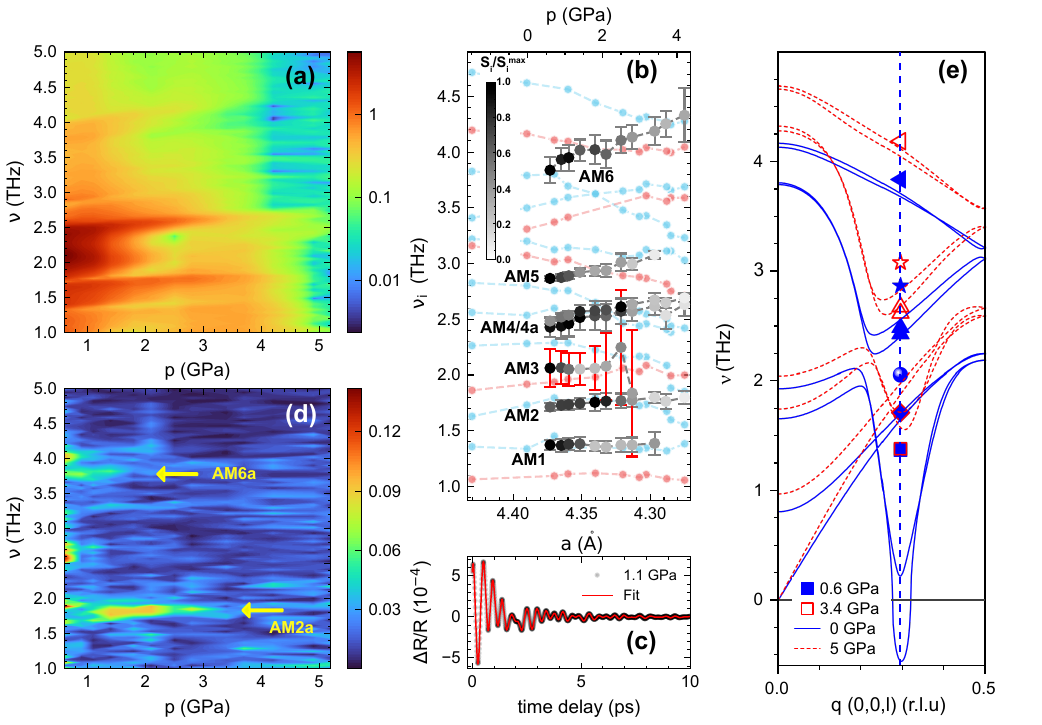}

\caption{\textbf{Pressure-dependence of amplitude modes in CeTe$_{3}$ at room
temperature.} (a) The p-dependence of the FFT intensity spectrum (note the
logarithmic scale). (b) Mode frequencies, $\nu_{i}$, normalized spectral
weights, S$_{i}/$S$_{i}^{\max}$, and dampings, $\Gamma_{i}$ (given by
the error-bars) extracted from the real-time fit of the oscillatory response. For
comparison with chemical pressure we include the data adapted from the recent
systematic spontaneous Raman study.\cite{yumigeta2022} Here, light-red
symbols correspond to Raman active modes of the high temperature normal phase
(present both above and below T$_{CDW}$), while light-blue circles denote the
frequencies of Raman active modes that appear in the CDW phase (which we refer
to as amplitude modes - see discussion above). Raman data were recorded at 79
K. The relation between the lattice constant and applied pressure follows Ref. [\onlinecite{sacchetti2009}]. (c) The oscillatory response at 1.1 GPa together with the real-time fit.
(d) The p-dependence of the FFT intensity spectrum of the residual (real-time
fit subtracted from the data) suggesting two additional weak modes at
$\approx1.8$ THz (AM2a) and $\approx3.7$ THz (AM6a). (e) Calculated dispersions of in-plane transversely polarized phonon modes along the [001] direction for ambient pressure (solid blue lines) and p = 5 GPa (dashed red lines), with vertical dashed blue line denoting the incommensurate CDW wavevector (0,0,$\approx2/7 c^{\ast}$). Frequencies of AMs recorded at 0.6 (solid blue symbols) and 3.4 GPa (open red symbols) at room temperature are added for comparison.}%
\label{Fig5}%
\end{figure}

Figure \ref{Fig5} presents the pressure dependence of the oscillatory response
in CeTe$_{3}$ at room temperature. The data are analysed by subtracting the
overdamped response from the raw data and evaluating the oscillatory response
in both frequency and time domain. Figure\ref{Fig5} (a) presents the
p-dependence of the FFT intensity spectrum, confirming the existence of
several amplitude modes. Their amplitudes gradually decrease with pressure,
eventually disappearing at pressures beyond $\approx5$ GPa. A more detailed
view of the modes' parameters is obtained by real-time fitting, an example of
such is shown in panel (c). Figure \ref{Fig5}(b) presents the extracted mode
frequencies, $\nu_{i}$, normalized spectral weights (S$_{i}/$S$_{i}^{\max}$,
where S$_{i}^{\max}$ is the maximal spectral weight of $i$-th mode) and dampings,
$\Gamma_{i}$ (given by error-bars). Using real-time fit we can distinguish two near-lying
modes at $\approx2.5$ THz (AM4 and AM4a). Finally, by subtracting the fits from the raw data
and plotting the FFT intensity spectrum of the residual as a function of
pressure, see Figure \ref{Fig5}(d), reveals the presence of two additional weak
modes\ at $\approx1.8$ THz and $\approx3.7$ THz.

In Figure \ref{Fig5}(b) we also include frequencies of Raman active modes as a
function of chemical pressure recorded at 79 K, adapted from the recent
systematic spontaneous Raman study.\cite{yumigeta2022} Here, Raman active
modes observed in both the normal and the CDW state are given by red circles,
while the modes appearing only in the low temperature CDW phase are
represented by solid blue circles. Considering the difference in temperatures
(79 K vs. 300 K) there is an overall very good correspondence between the
modes observed in pump-probe spectroscopy and the CDW-induced modes recorded
with spontaneous Raman scattering. Note that the intensities of the
CDW-induced modes are in spontaneous Raman scattering typically much lower
than intensities of the normal Raman active phonons, while in pump-probe
measurements probing coherent modes the situation is reversed. This can be
linked to the nature of the generation mechanism. In pump-probe studies in
opaque materials coherent excitation is displacive in nature,\cite{zeiger1992} thus
the modes that are strongly coupled to the electronic system are excited.
Indeed, the CDW amplitude modes present a prime example of such strongly
coupled modes, where perturbation of electronic subsystem directly affects the
equilibrium atomic positions, launching coherent dynamics of the order
parameter including both electronic and lattice degrees of freedom.

It follows from Figure \ref{Fig5}(b) that most of the amplitude modes harden
with increasing pressure, with the slopes ranging from $\approx150$ GHz/GPa
for the highest frequency mode (AM6), to nearly zero for the lowest frequency
amplitude mode (AM1). Even the dominant $\approx2$ THz mode (AM3), with the strongest
coupling to electronic order, reflected in the highest oscillator strength and
damping, shows only weak softening as a function of pressure, before becoming
overdamped near 3 GPa. The observed hardening of modes can be naturally accounted for
by considering the nature of amplitude modes, being a result of coupling of
electronic order to bare phonon modes at the q$_{CDW}$. Here, increasing
pressure should lead to hardening of the bare phonons competing with the mode softening as a result of pressure-driven reduction of the electronic order parameter $\Delta$. 
Figure \ref{Fig5}(e) presents the calculated phonon dispersions of in-plane transversely
polarized phonon modes along the [001] direction for ambient pressure and p = 5 GPa (see Methods). There is an overall good correspondence between the observed amplitude modes AM3-AM5 and normal phonons at (0,0,$\approx2/7 c^{\ast}$) both in terms of frequencies and the observed pressure-induced hardening. On the other hand, the weak low frequency modes AM1 and AM2, which display weak hardening with pressure and lack of hybridization with AM3, are likely a result of coupling to low-frequency acoustic modes at (0,0,$\approx2/7 c^{\ast}$) with longitudinal (AM2) and out-of-plane transverse (AM1) character - see Figure S5 of the SI.

\section*{Discussion}
We performed systematic temperature and pressure dependent studies of coherent order parameter dynamics in a prototypical CDW system CeTe$_{3}$. In the CDW state, we observe several amplitude modes, which can be attributed to coupling of the electronic order parameter to phonons at the CDW wavevector. From the temperature and pressure dependence of modes' parameters (Figs. \ref{Fig2} and \ref{Fig5}) and comparison to phonon dispersion calculations (Figure \ref{Fig5}(e)) we confirm that coupling to transverse in-plane polarized optical phonons governs the formation of CDW order, yet also coupling to the longitudinal and out-of-plane transverse acoustic modes at the CDW wavevector is substantial.

In the range of applied pressures up to $\approx 6$ GPa pressure-tuning results in a gradual suppression of the CDW transition temperature from $\approx 570$ K at ambient pressure to near room temperature, as shown by blue circles in Figure \ref{Fig1}(b). The effect of hydrostatic pressure on the CDW transition temperature is qualitatively similar to chemical substitution, with T$_{CDW}^{q_{1}}$(p) showing linear dependence on the lattice constant. However, the extracted T$_{CDW}^{q_{1}}$(p) are consistently higher than in the case of chemical substitution. Moreover, no signatures of the second CDW transition, T$_{CDW}^{q_{2}}$, are observed with pressure-tuning. We attribute the observed fine differences in the CDW orders under hydrostatic and chemical pressures to the details in the electron-phonon coupling. Indeed, the frequencies of both normal state phonons and amplitude modes show different evolutions with hydrostatic and chemical pressure, see Figure \ref{Fig5}(b). While increase in hydrostatic pressure results in hardening of all modes, both hardening and softening of modes are observed with chemical pressure.\cite{yumigeta2022} 

Apart from the fine differences in the manipulation of the CDW order with hydrostatic and chemical pressure, experiments with pressure-tuning provide access to the pressure-dependence of the overall coupling strength of carriers to relevant normal modes at the CDW wavevector, which governs the damping of the overdamped electronic mode. Surprisingly, Figure \ref{Fig4}(c) reveals a pronounced increase in the electronic damping rate $\tau_{0}^{-1}$ with applied pressure, implying an increase in the overall electron-phonon coupling strength with pressure despite the concomitant reduction in T$_{CDW}^{q_{1}}$ (the latter also being proportional to the CDW gap magnitude.\cite{sacchetti2009,Zocco2015}) At face value, this observation suggests that the pressure-driven suppression of the CDW order is mainly governed by the pressure-induced changes in the Fermi-surface topology, i.e., reduced Fermi-surface nesting. This is in line with the early studies of evolution of the electronic structure in RTe$_{3}$ as a function of ionic radius of R$^{3+}$, which revealed that decreasing the ionic radius results in a decrease in the area of gapped Fermi surface in the CDW phase.\cite{ARPES2008} On the other hand, there is mounting evidence that momentum-dependent electron-phonon coupling governs the CDW wavevector in prototypical two-dimensional CDW systems such as RTe$_3$.\cite{mazin2008,eiter2012,maschek2015,maschek2018} Furthermore, our phonon dispersion calculations demonstrate hardening of the bare phonons with pressure (Figure S5 of the SI) which may compete with an increase in the electron-phonon coupling strength. 

The observed anti-correlation between electron-phonon coupling strength and T$_{CDW}$ ($\Delta_{CDW}$) therefore demonstrates that the stability of the CDW in pressure-tuned CeTe$_3$ is governed by the subtle interplay between the Fermi surface nesting and momentum-dependent electron-phonon coupling, providing new insight into the microscopic mechanism of CDW formation in this canonical two-dimensional system.

For pressures beyond $\sim 6$ GPa no evidence of CDW order is observed down to cryogenic temperatures. Moreover, in this pressure range the electronic relaxation exhibits a pronounced slowing down with decreasing temperature, as shown in Figure \ref{Fig3}(b2). Such a slowing down is not observed in metals, where the relaxation time monotonically increases with increasing temperature. \cite{groeneveld1995,Ahn2004} Instead, it is commonly observed in heavy electron systems, where interaction between localized f-moments and conduction electrons leads to hybridization gaps and relaxation bottlenecks.\cite{demsar2006a,demsar2006b} Since in CeTe$_{3}$ renormalization of itinerant bands in the vicinity of the Ce 4f multiplet has been observed already at ambient pressure,\cite{ARPES2008,Saunot2026} we attribute the observed anomalous slowing down of carrier relaxation at low temperatures to pressure-driven enhancement of hybridization between localized Ce 4f levels and itinerant carriers suggestive of emerging heavy-electron behavior.

In summary, our real-time spectroscopic studies of the evolution of the macroscopic order in CeTe$_{3}$ with pressure provide important insights into the nature of the CDW ground state and the corresponding collective modes in this canonical CDW system. Together with the observed signatures of the emergent heavy-electron behavior at high pressures the time-resolved approach paves the way to future studies of quantum materials with competing orders and the quest for novel metastable states. 

\bigskip

\section*{Methods}

We studied the pressure (p) and temperature (T) dependence of the photoinduced reflectivity dynamics in CeTe$_{3}$ single crystals 
using optical pump-probe technique.
\subsection*{Femtosecond time-resolved spectroscopy}
The experiments were performed using a Ti:Sapphire laser amplifier operating at
a repetition rate of 250 kHz, pulse duration of 60 fs at a central wavelength
of 800 nm ($\lambda_{0}$) and bandwidth of 25 nm ($\Delta\lambda_{0}$). For
photoexcitation we used both near-infrared (800 nm) and frequency doubled (400
nm) pulses, while probing changes in reflectivity in the near-infrared (800 nm).
Pump/probe pulses were focused to full-width-at-half-maximum (FWHM) spot sizes
of 40 $\mu$m and 25 $\mu$m, respectively, at near-normal incidence.

$\Delta$R/R(t) traces were measured using fast-scan technique with
differential detection and modulation of the pump pulse at half the repetition
rate using an electro optic modulator (EOM), providing high dynamic range.\cite{schaefer2010,warawa2023}

\subsection*{Samples}
Single crystals of CeTe$_{3}$ were synthesized via a self-flux method.\cite{ru2006,ru2008} Typical crystals had a surface area around 1 mm$^2$ and were 0.5-mm thick. For high-pressure studies, the crystals were cut and thinned down by mechanical
exfoliation and placed into the 200 $\mu$m culet of the
diamond anvil cell (DAC), as depicted in the inset to Figure 1(b). 

\subsection*{High pressure cell}
We used CryoDAC-ST (Almax-easyLab) pressure cell. The lateral dimensions of crystals in the DAC cell were of the order of 100-120 $\mu$m. We used cesium iodide salt (CsI) as a pressure medium. The pressure was recorded via fluorescence line shifts of Ruby crystals put next to the sample in the pressure medium.\cite{forman1972pressure} After loading the sample, the DAC was placed in a continuous flow cryostat (Janis, ST-500). Temperature dependent measurements were performed upon warming, starting at 10 K.

\subsection*{Phonon dispersion calculations}

Lattice dynamics properties for the high-temperature orthorombic structure
were calculated using the linear response or density-functional perturbation
theory (DFPT) in the framework of the mixed basis pseudopotential
method.\cite{heid1999} To avoid complications with the f-electrons in the standard density-functional theory, we have performed our calculations for LaTe$_3$, with the atomic mass of La substituted
by that of Ce, and using lattice constants from Ref.[\onlinecite{sacchetti2009}]. Further computational details can be found in Refs. \onlinecite{maschek2015,maschek2018}.

\section*{Data availability}

All data supporting the findings of this study are available from the corresponding author upon reasonable request.

\begin{acknowledgments}
This work was funded by the German Research Foundation (DFG, Deutsche Forschungsgemeinschaft), grant no. TRR288 422213477 (project B08) and TRR173 268565370 (project A05). R.H. acknowledges support by the state of Baden-Württemberg through bwHPC.
\end{acknowledgments}

\section*{Author Contributions}

A.P. and J.D. conceived the project. P.G-G. and I.R.F. grew and characterized CeTe$_3$ single crystals. 
C.V.K., P.Y., A.P.L., J.T., M.O., H.S. performed experiments. C.V.K., P.Y. and J.D. analysed the data and prepared figures. R.H. performed phonon dispersion calculations.
J.D. wrote the manuscript with input from all authors. All authors participated in discussing the results and commenting on the article.

\section*{Competing interests}

The authors declare no competing interests.


\begin{thebibliography}{79}%
\makeatletter
\providecommand \@ifxundefined [1]{%
 \@ifx{#1\undefined}
}%
\providecommand \@ifnum [1]{%
 \ifnum #1\expandafter \@firstoftwo
 \else \expandafter \@secondoftwo
 \fi
}%
\providecommand \@ifx [1]{%
 \ifx #1\expandafter \@firstoftwo
 \else \expandafter \@secondoftwo
 \fi
}%
\providecommand \natexlab [1]{#1}%
\providecommand \enquote  [1]{``#1''}%
\providecommand \bibnamefont  [1]{#1}%
\providecommand \bibfnamefont [1]{#1}%
\providecommand \citenamefont [1]{#1}%
\providecommand \href@noop [0]{\@secondoftwo}%
\providecommand \href [0]{\begingroup \@sanitize@url \@href}%
\providecommand \@href[1]{\@@startlink{#1}\@@href}%
\providecommand \@@href[1]{\endgroup#1\@@endlink}%
\providecommand \@sanitize@url [0]{\catcode `\\12\catcode `\$12\catcode
  `\&12\catcode `\#12\catcode `\^12\catcode `\_12\catcode `\%12\relax}%
\providecommand \@@startlink[1]{}%
\providecommand \@@endlink[0]{}%
\providecommand \url  [0]{\begingroup\@sanitize@url \@url }%
\providecommand \@url [1]{\endgroup\@href {#1}{\urlprefix }}%
\providecommand \urlprefix  [0]{URL }%
\providecommand \Eprint [0]{\href }%
\providecommand \doibase [0]{https://doi.org/}%
\providecommand \selectlanguage [0]{\@gobble}%
\providecommand \bibinfo  [0]{\@secondoftwo}%
\providecommand \bibfield  [0]{\@secondoftwo}%
\providecommand \translation [1]{[#1]}%
\providecommand \BibitemOpen [0]{}%
\providecommand \bibitemStop [0]{}%
\providecommand \bibitemNoStop [0]{.\EOS\space}%
\providecommand \EOS [0]{\spacefactor3000\relax}%
\providecommand \BibitemShut  [1]{\csname bibitem#1\endcsname}%
\let\auto@bib@innerbib\@empty
\bibitem [{\citenamefont {Gr{\"u}ner}(1988)}]{gruner1988}%
  \BibitemOpen
  \bibfield  {author} {\bibinfo {author} {\bibfnamefont {G.}~\bibnamefont
  {Gr{\"u}ner}},\ }\bibfield  {title} {\enquote {\bibinfo {title} {The dynamics
  of charge-density waves},}\ }\href@noop {} {\bibfield  {journal} {\bibinfo
  {journal} {Reviews of modern physics}\ }\textbf {\bibinfo {volume} {60}},\
  \bibinfo {pages} {1129} (\bibinfo {year} {1988})}\BibitemShut {NoStop}%
\bibitem [{\citenamefont {Monceau}(2012)}]{Monceau2012}%
  \BibitemOpen
  \bibfield  {author} {\bibinfo {author} {\bibfnamefont {P.}~\bibnamefont
  {Monceau}},\ }\bibfield  {title} {\enquote {\bibinfo {title} {Electronic
  crystals: an experimental overview},}\ }\href
  {https://doi.org/10.1080/00018732.2012.719674} {\bibfield  {journal}
  {\bibinfo  {journal} {Advances in Physics}\ }\textbf {\bibinfo {volume}
  {61}},\ \bibinfo {pages} {325--581} (\bibinfo {year} {2012})}\BibitemShut
  {NoStop}%
\bibitem [{\citenamefont {Wise}\ \emph {et~al.}(2008)\citenamefont {Wise},
  \citenamefont {Boyer}, \citenamefont {Chatterjee}, \citenamefont {Kondo},
  \citenamefont {Takeuchi}, \citenamefont {Ikuta}, \citenamefont {Wang},\ and\
  \citenamefont {Hudson}}]{wise2008}%
  \BibitemOpen
  \bibfield  {author} {\bibinfo {author} {\bibfnamefont {W.}~\bibnamefont
  {Wise}}, \bibinfo {author} {\bibfnamefont {M.}~\bibnamefont {Boyer}},
  \bibinfo {author} {\bibfnamefont {K.}~\bibnamefont {Chatterjee}}, \bibinfo
  {author} {\bibfnamefont {T.}~\bibnamefont {Kondo}}, \bibinfo {author}
  {\bibfnamefont {T.}~\bibnamefont {Takeuchi}}, \bibinfo {author}
  {\bibfnamefont {H.}~\bibnamefont {Ikuta}}, \bibinfo {author} {\bibfnamefont
  {Y.}~\bibnamefont {Wang}},\ and\ \bibinfo {author} {\bibfnamefont
  {E.}~\bibnamefont {Hudson}},\ }\bibfield  {title} {\enquote {\bibinfo {title}
  {Charge-density-wave origin of cuprate checkerboard visualized by scanning
  tunnelling microscopy},}\ }\href@noop {} {\bibfield  {journal} {\bibinfo
  {journal} {Nature physics}\ }\textbf {\bibinfo {volume} {4}},\ \bibinfo
  {pages} {696--699} (\bibinfo {year} {2008})}\BibitemShut {NoStop}%
\bibitem [{\citenamefont {Ghiringhelli}\ \emph {et~al.}(2012)\citenamefont
  {Ghiringhelli}, \citenamefont {Le~Tacon}, \citenamefont {Minola},
  \citenamefont {Blanco-Canosa}, \citenamefont {Mazzoli}, \citenamefont
  {Brookes}, \citenamefont {De~Luca}, \citenamefont {Frano}, \citenamefont
  {Hawthorn}, \citenamefont {He} \emph {et~al.}}]{Ghiringhelli2012}%
  \BibitemOpen
  \bibfield  {author} {\bibinfo {author} {\bibfnamefont {G.}~\bibnamefont
  {Ghiringhelli}}, \bibinfo {author} {\bibfnamefont {M.}~\bibnamefont
  {Le~Tacon}}, \bibinfo {author} {\bibfnamefont {M.}~\bibnamefont {Minola}},
  \bibinfo {author} {\bibfnamefont {S.}~\bibnamefont {Blanco-Canosa}}, \bibinfo
  {author} {\bibfnamefont {C.}~\bibnamefont {Mazzoli}}, \bibinfo {author}
  {\bibfnamefont {N.}~\bibnamefont {Brookes}}, \bibinfo {author} {\bibfnamefont
  {G.}~\bibnamefont {De~Luca}}, \bibinfo {author} {\bibfnamefont
  {A.}~\bibnamefont {Frano}}, \bibinfo {author} {\bibfnamefont
  {D.}~\bibnamefont {Hawthorn}}, \bibinfo {author} {\bibfnamefont
  {F.}~\bibnamefont {He}}, \emph {et~al.},\ }\bibfield  {title} {\enquote
  {\bibinfo {title} {{Long-range incommensurate charge fluctuations in (Y, Nd)
  Ba2Cu3O6+x}},}\ }\href {https://doi.org/10.1126/science.1223532} {\bibfield
  {journal} {\bibinfo  {journal} {Science}\ }\textbf {\bibinfo {volume}
  {337}},\ \bibinfo {pages} {821--825} (\bibinfo {year} {2012})}\BibitemShut
  {NoStop}%
\bibitem [{\citenamefont {Torchinsky}\ \emph {et~al.}(2013)\citenamefont
  {Torchinsky}, \citenamefont {Mahmood}, \citenamefont {Bollinger},
  \citenamefont {Bo{\v{z}}ovi{\'c}},\ and\ \citenamefont
  {Gedik}}]{torchinsky2013}%
  \BibitemOpen
  \bibfield  {author} {\bibinfo {author} {\bibfnamefont {D.~H.}\ \bibnamefont
  {Torchinsky}}, \bibinfo {author} {\bibfnamefont {F.}~\bibnamefont {Mahmood}},
  \bibinfo {author} {\bibfnamefont {A.~T.}\ \bibnamefont {Bollinger}}, \bibinfo
  {author} {\bibfnamefont {I.}~\bibnamefont {Bo{\v{z}}ovi{\'c}}},\ and\
  \bibinfo {author} {\bibfnamefont {N.}~\bibnamefont {Gedik}},\ }\bibfield
  {title} {\enquote {\bibinfo {title} {Fluctuating charge-density waves in a
  cuprate superconductor},}\ }\href@noop {} {\bibfield  {journal} {\bibinfo
  {journal} {Nature materials}\ }\textbf {\bibinfo {volume} {12}},\ \bibinfo
  {pages} {387--391} (\bibinfo {year} {2013})}\BibitemShut {NoStop}%
\bibitem [{\citenamefont {da~Silva~Neto}\ \emph {et~al.}(2014)\citenamefont
  {da~Silva~Neto}, \citenamefont {Aynajian}, \citenamefont {Frano},
  \citenamefont {Comin}, \citenamefont {Schierle}, \citenamefont {Weschke},
  \citenamefont {Gyenis}, \citenamefont {Wen}, \citenamefont {Schneeloch},
  \citenamefont {Xu}, \citenamefont {Ono}, \citenamefont {Gu}, \citenamefont
  {Tacon},\ and\ \citenamefont {Yazdani}}]{daSilva2014}%
  \BibitemOpen
  \bibfield  {author} {\bibinfo {author} {\bibfnamefont {E.~H.}\ \bibnamefont
  {da~Silva~Neto}}, \bibinfo {author} {\bibfnamefont {P.}~\bibnamefont
  {Aynajian}}, \bibinfo {author} {\bibfnamefont {A.}~\bibnamefont {Frano}},
  \bibinfo {author} {\bibfnamefont {R.}~\bibnamefont {Comin}}, \bibinfo
  {author} {\bibfnamefont {E.}~\bibnamefont {Schierle}}, \bibinfo {author}
  {\bibfnamefont {E.}~\bibnamefont {Weschke}}, \bibinfo {author} {\bibfnamefont
  {A.}~\bibnamefont {Gyenis}}, \bibinfo {author} {\bibfnamefont
  {J.}~\bibnamefont {Wen}}, \bibinfo {author} {\bibfnamefont {J.}~\bibnamefont
  {Schneeloch}}, \bibinfo {author} {\bibfnamefont {Z.}~\bibnamefont {Xu}},
  \bibinfo {author} {\bibfnamefont {S.}~\bibnamefont {Ono}}, \bibinfo {author}
  {\bibfnamefont {G.}~\bibnamefont {Gu}}, \bibinfo {author} {\bibfnamefont
  {M.~L.}\ \bibnamefont {Tacon}},\ and\ \bibinfo {author} {\bibfnamefont
  {A.}~\bibnamefont {Yazdani}},\ }\bibfield  {title} {\enquote {\bibinfo
  {title} {Ubiquitous interplay between charge ordering and high-temperature
  superconductivity in cuprates},}\ }\href
  {https://doi.org/10.1126/science.1243479} {\bibfield  {journal} {\bibinfo
  {journal} {Science}\ }\textbf {\bibinfo {volume} {343}},\ \bibinfo {pages}
  {393--396} (\bibinfo {year} {2014})}\BibitemShut {NoStop}%
\bibitem [{\citenamefont {Miao}\ \emph {et~al.}(2021)\citenamefont {Miao},
  \citenamefont {Fabbris}, \citenamefont {Koch}, \citenamefont {Mazzone},
  \citenamefont {Nelson}, \citenamefont {Acevedo-Esteves}, \citenamefont {Gu},
  \citenamefont {Li}, \citenamefont {Yilimaz}, \citenamefont {Kaznatcheev}
  \emph {et~al.}}]{miao2021}%
  \BibitemOpen
  \bibfield  {author} {\bibinfo {author} {\bibfnamefont {H.}~\bibnamefont
  {Miao}}, \bibinfo {author} {\bibfnamefont {G.}~\bibnamefont {Fabbris}},
  \bibinfo {author} {\bibfnamefont {R.}~\bibnamefont {Koch}}, \bibinfo {author}
  {\bibfnamefont {D.}~\bibnamefont {Mazzone}}, \bibinfo {author} {\bibfnamefont
  {C.}~\bibnamefont {Nelson}}, \bibinfo {author} {\bibfnamefont
  {R.}~\bibnamefont {Acevedo-Esteves}}, \bibinfo {author} {\bibfnamefont
  {G.}~\bibnamefont {Gu}}, \bibinfo {author} {\bibfnamefont {Y.}~\bibnamefont
  {Li}}, \bibinfo {author} {\bibfnamefont {T.}~\bibnamefont {Yilimaz}},
  \bibinfo {author} {\bibfnamefont {K.}~\bibnamefont {Kaznatcheev}}, \emph
  {et~al.},\ }\bibfield  {title} {\enquote {\bibinfo {title} {Charge density
  waves in cuprate superconductors beyond the critical doping},}\ }\href@noop
  {} {\bibfield  {journal} {\bibinfo  {journal} {npj Quantum Materials}\
  }\textbf {\bibinfo {volume} {6}},\ \bibinfo {pages} {31} (\bibinfo {year}
  {2021})}\BibitemShut {NoStop}%
\bibitem [{\citenamefont {Tam}\ \emph {et~al.}(2022)\citenamefont {Tam},
  \citenamefont {Choi}, \citenamefont {Ding}, \citenamefont {Agrestini},
  \citenamefont {Nag}, \citenamefont {Wu}, \citenamefont {Huang}, \citenamefont
  {Luo}, \citenamefont {Gao}, \citenamefont {Garc{\'\i}a-Fern{\'a}ndez} \emph
  {et~al.}}]{tam2022}%
  \BibitemOpen
  \bibfield  {author} {\bibinfo {author} {\bibfnamefont {C.~C.}\ \bibnamefont
  {Tam}}, \bibinfo {author} {\bibfnamefont {J.}~\bibnamefont {Choi}}, \bibinfo
  {author} {\bibfnamefont {X.}~\bibnamefont {Ding}}, \bibinfo {author}
  {\bibfnamefont {S.}~\bibnamefont {Agrestini}}, \bibinfo {author}
  {\bibfnamefont {A.}~\bibnamefont {Nag}}, \bibinfo {author} {\bibfnamefont
  {M.}~\bibnamefont {Wu}}, \bibinfo {author} {\bibfnamefont {B.}~\bibnamefont
  {Huang}}, \bibinfo {author} {\bibfnamefont {H.}~\bibnamefont {Luo}}, \bibinfo
  {author} {\bibfnamefont {P.}~\bibnamefont {Gao}}, \bibinfo {author}
  {\bibfnamefont {M.}~\bibnamefont {Garc{\'\i}a-Fern{\'a}ndez}}, \emph
  {et~al.},\ }\bibfield  {title} {\enquote {\bibinfo {title} {{Charge density
  waves in infinite-layer NdNiO2 nickelates}},}\ }\href@noop {} {\bibfield
  {journal} {\bibinfo  {journal} {Nature Materials}\ }\textbf {\bibinfo
  {volume} {21}},\ \bibinfo {pages} {1116--1120} (\bibinfo {year}
  {2022})}\BibitemShut {NoStop}%
\bibitem [{\citenamefont {Xu}\ \emph {et~al.}(2025)\citenamefont {Xu},
  \citenamefont {Wang}, \citenamefont {Huo}, \citenamefont {Hu}, \citenamefont
  {Wu}, \citenamefont {Yue}, \citenamefont {Wu}, \citenamefont {Wang},
  \citenamefont {Dong},\ and\ \citenamefont {Wang}}]{xu2025}%
  \BibitemOpen
  \bibfield  {author} {\bibinfo {author} {\bibfnamefont {S.}~\bibnamefont
  {Xu}}, \bibinfo {author} {\bibfnamefont {H.}~\bibnamefont {Wang}}, \bibinfo
  {author} {\bibfnamefont {M.}~\bibnamefont {Huo}}, \bibinfo {author}
  {\bibfnamefont {D.}~\bibnamefont {Hu}}, \bibinfo {author} {\bibfnamefont
  {Q.}~\bibnamefont {Wu}}, \bibinfo {author} {\bibfnamefont {L.}~\bibnamefont
  {Yue}}, \bibinfo {author} {\bibfnamefont {D.}~\bibnamefont {Wu}}, \bibinfo
  {author} {\bibfnamefont {M.}~\bibnamefont {Wang}}, \bibinfo {author}
  {\bibfnamefont {T.}~\bibnamefont {Dong}},\ and\ \bibinfo {author}
  {\bibfnamefont {N.}~\bibnamefont {Wang}},\ }\bibfield  {title} {\enquote
  {\bibinfo {title} {Collapse of density wave and emergence of
  superconductivity in pressurized-la4ni3o10 evidenced by ultrafast
  spectroscopy},}\ }\href@noop {} {\bibfield  {journal} {\bibinfo  {journal}
  {Nature Communications}\ }\textbf {\bibinfo {volume} {16}},\ \bibinfo {pages}
  {7039} (\bibinfo {year} {2025})}\BibitemShut {NoStop}%
\bibitem [{\citenamefont {Souliou}\ \emph {et~al.}(2022)\citenamefont
  {Souliou}, \citenamefont {Lacmann}, \citenamefont {Heid}, \citenamefont
  {Meingast}, \citenamefont {Frachet}, \citenamefont {Paolasini}, \citenamefont
  {Haghighirad}, \citenamefont {Merz}, \citenamefont {Bosak},\ and\
  \citenamefont {Le~Tacon}}]{souliou2022}%
  \BibitemOpen
  \bibfield  {author} {\bibinfo {author} {\bibfnamefont {S.}~\bibnamefont
  {Souliou}}, \bibinfo {author} {\bibfnamefont {T.}~\bibnamefont {Lacmann}},
  \bibinfo {author} {\bibfnamefont {R.}~\bibnamefont {Heid}}, \bibinfo {author}
  {\bibfnamefont {C.}~\bibnamefont {Meingast}}, \bibinfo {author}
  {\bibfnamefont {M.}~\bibnamefont {Frachet}}, \bibinfo {author} {\bibfnamefont
  {L.}~\bibnamefont {Paolasini}}, \bibinfo {author} {\bibfnamefont {A.-A.}\
  \bibnamefont {Haghighirad}}, \bibinfo {author} {\bibfnamefont
  {M.}~\bibnamefont {Merz}}, \bibinfo {author} {\bibfnamefont {A.}~\bibnamefont
  {Bosak}},\ and\ \bibinfo {author} {\bibfnamefont {M.}~\bibnamefont
  {Le~Tacon}},\ }\bibfield  {title} {\enquote {\bibinfo {title} {{Soft-phonon
  and charge-density-wave formation in nematic BaNi2As2}},}\ }\href@noop {}
  {\bibfield  {journal} {\bibinfo  {journal} {Physical Review Letters}\
  }\textbf {\bibinfo {volume} {129}},\ \bibinfo {pages} {247602} (\bibinfo
  {year} {2022})}\BibitemShut {NoStop}%
\bibitem [{\citenamefont {Pokharel}\ \emph {et~al.}(2022)\citenamefont
  {Pokharel}, \citenamefont {Grigorev}, \citenamefont {Mejas}, \citenamefont
  {Dong}, \citenamefont {Haghighirad}, \citenamefont {Heid}, \citenamefont
  {Yao}, \citenamefont {Merz}, \citenamefont {Le~Tacon},\ and\ \citenamefont
  {Demsar}}]{pokharel2022}%
  \BibitemOpen
  \bibfield  {author} {\bibinfo {author} {\bibfnamefont {A.~R.}\ \bibnamefont
  {Pokharel}}, \bibinfo {author} {\bibfnamefont {V.}~\bibnamefont {Grigorev}},
  \bibinfo {author} {\bibfnamefont {A.}~\bibnamefont {Mejas}}, \bibinfo
  {author} {\bibfnamefont {T.}~\bibnamefont {Dong}}, \bibinfo {author}
  {\bibfnamefont {A.~A.}\ \bibnamefont {Haghighirad}}, \bibinfo {author}
  {\bibfnamefont {R.}~\bibnamefont {Heid}}, \bibinfo {author} {\bibfnamefont
  {Y.}~\bibnamefont {Yao}}, \bibinfo {author} {\bibfnamefont {M.}~\bibnamefont
  {Merz}}, \bibinfo {author} {\bibfnamefont {M.}~\bibnamefont {Le~Tacon}},\
  and\ \bibinfo {author} {\bibfnamefont {J.}~\bibnamefont {Demsar}},\
  }\bibfield  {title} {\enquote {\bibinfo {title} {{Dynamics of collective
  modes in an unconventional charge density wave system BaNi2As2}},}\
  }\href@noop {} {\bibfield  {journal} {\bibinfo  {journal} {Communications
  Physics}\ }\textbf {\bibinfo {volume} {5}},\ \bibinfo {pages} {141} (\bibinfo
  {year} {2022})}\BibitemShut {NoStop}%
\bibitem [{\citenamefont {Kolincio}\ \emph {et~al.}(2017)\citenamefont
  {Kolincio}, \citenamefont {Roman}, \citenamefont {Winiarski}, \citenamefont
  {Strychalska-Nowak},\ and\ \citenamefont {Klimczuk}}]{kolincio2017}%
  \BibitemOpen
  \bibfield  {author} {\bibinfo {author} {\bibfnamefont {K.~K.}\ \bibnamefont
  {Kolincio}}, \bibinfo {author} {\bibfnamefont {M.}~\bibnamefont {Roman}},
  \bibinfo {author} {\bibfnamefont {M.~J.}\ \bibnamefont {Winiarski}}, \bibinfo
  {author} {\bibfnamefont {J.}~\bibnamefont {Strychalska-Nowak}},\ and\
  \bibinfo {author} {\bibfnamefont {T.}~\bibnamefont {Klimczuk}},\ }\bibfield
  {title} {\enquote {\bibinfo {title} {{Magnetism and charge density waves in R
  NiC 2 (R= Ce, Pr, Nd)}},}\ }\href@noop {} {\bibfield  {journal} {\bibinfo
  {journal} {Physical Review B}\ }\textbf {\bibinfo {volume} {95}},\ \bibinfo
  {pages} {235156} (\bibinfo {year} {2017})}\BibitemShut {NoStop}%
\bibitem [{\citenamefont {Teng}\ \emph {et~al.}(2023)\citenamefont {Teng},
  \citenamefont {Oh}, \citenamefont {Tan}, \citenamefont {Chen}, \citenamefont
  {Huang}, \citenamefont {Gao}, \citenamefont {Yin}, \citenamefont {Chu},
  \citenamefont {Hashimoto}, \citenamefont {Lu} \emph {et~al.}}]{teng2023}%
  \BibitemOpen
  \bibfield  {author} {\bibinfo {author} {\bibfnamefont {X.}~\bibnamefont
  {Teng}}, \bibinfo {author} {\bibfnamefont {J.~S.}\ \bibnamefont {Oh}},
  \bibinfo {author} {\bibfnamefont {H.}~\bibnamefont {Tan}}, \bibinfo {author}
  {\bibfnamefont {L.}~\bibnamefont {Chen}}, \bibinfo {author} {\bibfnamefont
  {J.}~\bibnamefont {Huang}}, \bibinfo {author} {\bibfnamefont
  {B.}~\bibnamefont {Gao}}, \bibinfo {author} {\bibfnamefont {J.-X.}\
  \bibnamefont {Yin}}, \bibinfo {author} {\bibfnamefont {J.-H.}\ \bibnamefont
  {Chu}}, \bibinfo {author} {\bibfnamefont {M.}~\bibnamefont {Hashimoto}},
  \bibinfo {author} {\bibfnamefont {D.}~\bibnamefont {Lu}}, \emph {et~al.},\
  }\bibfield  {title} {\enquote {\bibinfo {title} {{Magnetism and charge
  density wave order in kagome FeGe}},}\ }\href@noop {} {\bibfield  {journal}
  {\bibinfo  {journal} {Nature physics}\ }\textbf {\bibinfo {volume} {19}},\
  \bibinfo {pages} {814--822} (\bibinfo {year} {2023})}\BibitemShut {NoStop}%
\bibitem [{\citenamefont {von Ungern-Sternberg~Schwark}\ \emph
  {et~al.}(2026)\citenamefont {von Ungern-Sternberg~Schwark}, \citenamefont
  {Haghighirad}, \citenamefont {Heid}, \citenamefont {McGuinness},
  \citenamefont {Maraytta}, \citenamefont {Eich}, \citenamefont {Merz},
  \citenamefont {Bosak}, \citenamefont {Chaney}, \citenamefont {Chumakova},
  \citenamefont {Pawbake}, \citenamefont {Faugeras}, \citenamefont {Tacon},\
  and\ \citenamefont {Souliou}}]{schwark2026}%
  \BibitemOpen
  \bibfield  {author} {\bibinfo {author} {\bibfnamefont {A.}~\bibnamefont {von
  Ungern-Sternberg~Schwark}}, \bibinfo {author} {\bibfnamefont {A.~A.}\
  \bibnamefont {Haghighirad}}, \bibinfo {author} {\bibfnamefont
  {R.}~\bibnamefont {Heid}}, \bibinfo {author} {\bibfnamefont {P.~H.}\
  \bibnamefont {McGuinness}}, \bibinfo {author} {\bibfnamefont
  {N.}~\bibnamefont {Maraytta}}, \bibinfo {author} {\bibfnamefont
  {A.}~\bibnamefont {Eich}}, \bibinfo {author} {\bibfnamefont {M.}~\bibnamefont
  {Merz}}, \bibinfo {author} {\bibfnamefont {A.}~\bibnamefont {Bosak}},
  \bibinfo {author} {\bibfnamefont {D.~A.}\ \bibnamefont {Chaney}}, \bibinfo
  {author} {\bibfnamefont {A.}~\bibnamefont {Chumakova}}, \bibinfo {author}
  {\bibfnamefont {A.}~\bibnamefont {Pawbake}}, \bibinfo {author} {\bibfnamefont
  {C.}~\bibnamefont {Faugeras}}, \bibinfo {author} {\bibfnamefont {M.~L.}\
  \bibnamefont {Tacon}},\ and\ \bibinfo {author} {\bibfnamefont {S.~M.}\
  \bibnamefont {Souliou}},\ }\href {https://arxiv.org/abs/2601.07255} {\enquote
  {\bibinfo {title} {{Interplay of Charge and Magnetic Orders in SmNiC$_2$
  Mediated by Electron-Phonon Interaction}},}\ } (\bibinfo {year} {2026}),\
  \Eprint {https://arxiv.org/abs/2601.07255} {arXiv:2601.07255} \BibitemShut
  {NoStop}%
\bibitem [{\citenamefont {Tan}\ \emph {et~al.}(2021)\citenamefont {Tan},
  \citenamefont {Liu}, \citenamefont {Wang},\ and\ \citenamefont
  {Yan}}]{tan2021}%
  \BibitemOpen
  \bibfield  {author} {\bibinfo {author} {\bibfnamefont {H.}~\bibnamefont
  {Tan}}, \bibinfo {author} {\bibfnamefont {Y.}~\bibnamefont {Liu}}, \bibinfo
  {author} {\bibfnamefont {Z.}~\bibnamefont {Wang}},\ and\ \bibinfo {author}
  {\bibfnamefont {B.}~\bibnamefont {Yan}},\ }\bibfield  {title} {\enquote
  {\bibinfo {title} {Charge density waves and electronic properties of
  superconducting kagome metals},}\ }\href@noop {} {\bibfield  {journal}
  {\bibinfo  {journal} {Physical review letters}\ }\textbf {\bibinfo {volume}
  {127}},\ \bibinfo {pages} {046401} (\bibinfo {year} {2021})}\BibitemShut
  {NoStop}%
\bibitem [{\citenamefont {Wang}\ \emph {et~al.}(2022)\citenamefont {Wang},
  \citenamefont {Petrides}, \citenamefont {McNamara}, \citenamefont {Hosen},
  \citenamefont {Lei}, \citenamefont {Wu}, \citenamefont {Hart}, \citenamefont
  {Lv}, \citenamefont {Yan}, \citenamefont {Xiao} \emph {et~al.}}]{wang2022}%
  \BibitemOpen
  \bibfield  {author} {\bibinfo {author} {\bibfnamefont {Y.}~\bibnamefont
  {Wang}}, \bibinfo {author} {\bibfnamefont {I.}~\bibnamefont {Petrides}},
  \bibinfo {author} {\bibfnamefont {G.}~\bibnamefont {McNamara}}, \bibinfo
  {author} {\bibfnamefont {M.~M.}\ \bibnamefont {Hosen}}, \bibinfo {author}
  {\bibfnamefont {S.}~\bibnamefont {Lei}}, \bibinfo {author} {\bibfnamefont
  {Y.-C.}\ \bibnamefont {Wu}}, \bibinfo {author} {\bibfnamefont {J.~L.}\
  \bibnamefont {Hart}}, \bibinfo {author} {\bibfnamefont {H.}~\bibnamefont
  {Lv}}, \bibinfo {author} {\bibfnamefont {J.}~\bibnamefont {Yan}}, \bibinfo
  {author} {\bibfnamefont {D.}~\bibnamefont {Xiao}}, \emph {et~al.},\
  }\bibfield  {title} {\enquote {\bibinfo {title} {Axial higgs mode detected by
  quantum pathway interference in rte3},}\ }\href@noop {} {\bibfield  {journal}
  {\bibinfo  {journal} {Nature}\ }\textbf {\bibinfo {volume} {606}},\ \bibinfo
  {pages} {896--901} (\bibinfo {year} {2022})}\BibitemShut {NoStop}%
\bibitem [{\citenamefont {Cao}\ \emph {et~al.}(2023)\citenamefont {Cao},
  \citenamefont {Xu}, \citenamefont {Fukui}, \citenamefont {Manjo},
  \citenamefont {Dong}, \citenamefont {Shi}, \citenamefont {Liu}, \citenamefont
  {Cao},\ and\ \citenamefont {Song}}]{cao2023}%
  \BibitemOpen
  \bibfield  {author} {\bibinfo {author} {\bibfnamefont {S.}~\bibnamefont
  {Cao}}, \bibinfo {author} {\bibfnamefont {C.}~\bibnamefont {Xu}}, \bibinfo
  {author} {\bibfnamefont {H.}~\bibnamefont {Fukui}}, \bibinfo {author}
  {\bibfnamefont {T.}~\bibnamefont {Manjo}}, \bibinfo {author} {\bibfnamefont
  {Y.}~\bibnamefont {Dong}}, \bibinfo {author} {\bibfnamefont {M.}~\bibnamefont
  {Shi}}, \bibinfo {author} {\bibfnamefont {Y.}~\bibnamefont {Liu}}, \bibinfo
  {author} {\bibfnamefont {C.}~\bibnamefont {Cao}},\ and\ \bibinfo {author}
  {\bibfnamefont {Y.}~\bibnamefont {Song}},\ }\bibfield  {title} {\enquote
  {\bibinfo {title} {{Competing charge-density wave instabilities in the kagome
  metal ScV6Sn6}},}\ }\href@noop {} {\bibfield  {journal} {\bibinfo  {journal}
  {Nature Communications}\ }\textbf {\bibinfo {volume} {14}},\ \bibinfo {pages}
  {7671} (\bibinfo {year} {2023})}\BibitemShut {NoStop}%
\bibitem [{\citenamefont {Elmers}\ \emph {et~al.}(2025)\citenamefont {Elmers},
  \citenamefont {Tkach}, \citenamefont {Lytvynenko}, \citenamefont {Yogi},
  \citenamefont {Schmitt}, \citenamefont {Biswas}, \citenamefont {Liu},
  \citenamefont {Chernov}, \citenamefont {Nguyen}, \citenamefont {Hoesch},
  \citenamefont {Kutnyakhov}, \citenamefont {Wind}, \citenamefont {Wenthaus},
  \citenamefont {Scholz}, \citenamefont {Rossnagel}, \citenamefont
  {Gloskovskii}, \citenamefont {Schlueter}, \citenamefont {Winkelmann},
  \citenamefont {Haghighirad}, \citenamefont {Lee}, \citenamefont {Sing},
  \citenamefont {Claessen}, \citenamefont {Le~Tacon}, \citenamefont {Demsar},
  \citenamefont {Sch\"onhense},\ and\ \citenamefont {Fedchenko}}]{Elmers2025}%
  \BibitemOpen
  \bibfield  {author} {\bibinfo {author} {\bibfnamefont {H.~J.}\ \bibnamefont
  {Elmers}}, \bibinfo {author} {\bibfnamefont {O.}~\bibnamefont {Tkach}},
  \bibinfo {author} {\bibfnamefont {Y.}~\bibnamefont {Lytvynenko}}, \bibinfo
  {author} {\bibfnamefont {P.}~\bibnamefont {Yogi}}, \bibinfo {author}
  {\bibfnamefont {M.}~\bibnamefont {Schmitt}}, \bibinfo {author} {\bibfnamefont
  {D.}~\bibnamefont {Biswas}}, \bibinfo {author} {\bibfnamefont
  {J.}~\bibnamefont {Liu}}, \bibinfo {author} {\bibfnamefont {S.~V.}\
  \bibnamefont {Chernov}}, \bibinfo {author} {\bibfnamefont {Q.}~\bibnamefont
  {Nguyen}}, \bibinfo {author} {\bibfnamefont {M.}~\bibnamefont {Hoesch}},
  \bibinfo {author} {\bibfnamefont {D.}~\bibnamefont {Kutnyakhov}}, \bibinfo
  {author} {\bibfnamefont {N.}~\bibnamefont {Wind}}, \bibinfo {author}
  {\bibfnamefont {L.}~\bibnamefont {Wenthaus}}, \bibinfo {author}
  {\bibfnamefont {M.}~\bibnamefont {Scholz}}, \bibinfo {author} {\bibfnamefont
  {K.}~\bibnamefont {Rossnagel}}, \bibinfo {author} {\bibfnamefont
  {A.}~\bibnamefont {Gloskovskii}}, \bibinfo {author} {\bibfnamefont
  {C.}~\bibnamefont {Schlueter}}, \bibinfo {author} {\bibfnamefont
  {A.}~\bibnamefont {Winkelmann}}, \bibinfo {author} {\bibfnamefont {A.-A.}\
  \bibnamefont {Haghighirad}}, \bibinfo {author} {\bibfnamefont {T.-L.}\
  \bibnamefont {Lee}}, \bibinfo {author} {\bibfnamefont {M.}~\bibnamefont
  {Sing}}, \bibinfo {author} {\bibfnamefont {R.}~\bibnamefont {Claessen}},
  \bibinfo {author} {\bibfnamefont {M.}~\bibnamefont {Le~Tacon}}, \bibinfo
  {author} {\bibfnamefont {J.}~\bibnamefont {Demsar}}, \bibinfo {author}
  {\bibfnamefont {G.}~\bibnamefont {Sch\"onhense}},\ and\ \bibinfo {author}
  {\bibfnamefont {O.}~\bibnamefont {Fedchenko}},\ }\bibfield  {title} {\enquote
  {\bibinfo {title} {{Chirality in the Kagome Metal
  ${\mathrm{CsV}}_{3}{\mathrm{Sb}}_{5}$}},}\ }\href
  {https://doi.org/10.1103/PhysRevLett.134.096401} {\bibfield  {journal}
  {\bibinfo  {journal} {Phys. Rev. Lett.}\ }\textbf {\bibinfo {volume} {134}},\
  \bibinfo {pages} {096401} (\bibinfo {year} {2025})}\BibitemShut {NoStop}%
\bibitem [{\citenamefont {Singh}\ \emph {et~al.}(2025)\citenamefont {Singh},
  \citenamefont {McNamara}, \citenamefont {Kim}, \citenamefont {Siddique},
  \citenamefont {Funni}, \citenamefont {Zhang}, \citenamefont {Luo},
  \citenamefont {Sakrikar}, \citenamefont {Kenney}, \citenamefont {Singha}
  \emph {et~al.}}]{singh2025}%
  \BibitemOpen
  \bibfield  {author} {\bibinfo {author} {\bibfnamefont {B.}~\bibnamefont
  {Singh}}, \bibinfo {author} {\bibfnamefont {G.}~\bibnamefont {McNamara}},
  \bibinfo {author} {\bibfnamefont {K.-M.}\ \bibnamefont {Kim}}, \bibinfo
  {author} {\bibfnamefont {S.}~\bibnamefont {Siddique}}, \bibinfo {author}
  {\bibfnamefont {S.~D.}\ \bibnamefont {Funni}}, \bibinfo {author}
  {\bibfnamefont {W.}~\bibnamefont {Zhang}}, \bibinfo {author} {\bibfnamefont
  {X.}~\bibnamefont {Luo}}, \bibinfo {author} {\bibfnamefont {P.}~\bibnamefont
  {Sakrikar}}, \bibinfo {author} {\bibfnamefont {E.~M.}\ \bibnamefont
  {Kenney}}, \bibinfo {author} {\bibfnamefont {R.}~\bibnamefont {Singha}},
  \emph {et~al.},\ }\bibfield  {title} {\enquote {\bibinfo {title} {Ferroaxial
  density wave from intertwined charge and orbital order in rare-earth
  tritellurides},}\ }\href@noop {} {\bibfield  {journal} {\bibinfo  {journal}
  {Nature Physics}\ }\textbf {\bibinfo {volume} {21}},\ \bibinfo {pages}
  {1578--1586} (\bibinfo {year} {2025})}\BibitemShut {NoStop}%
\bibitem [{\citenamefont {Yang}\ \emph {et~al.}(2022)\citenamefont {Yang},
  \citenamefont {He}, \citenamefont {Koo}, \citenamefont {Shen}, \citenamefont
  {Zhang}, \citenamefont {Liu}, \citenamefont {Liu}, \citenamefont {Chen},
  \citenamefont {Liang}, \citenamefont {Huang} \emph {et~al.}}]{yang2022}%
  \BibitemOpen
  \bibfield  {author} {\bibinfo {author} {\bibfnamefont {H.}~\bibnamefont
  {Yang}}, \bibinfo {author} {\bibfnamefont {K.}~\bibnamefont {He}}, \bibinfo
  {author} {\bibfnamefont {J.}~\bibnamefont {Koo}}, \bibinfo {author}
  {\bibfnamefont {S.}~\bibnamefont {Shen}}, \bibinfo {author} {\bibfnamefont
  {S.}~\bibnamefont {Zhang}}, \bibinfo {author} {\bibfnamefont
  {G.}~\bibnamefont {Liu}}, \bibinfo {author} {\bibfnamefont {Y.}~\bibnamefont
  {Liu}}, \bibinfo {author} {\bibfnamefont {C.}~\bibnamefont {Chen}}, \bibinfo
  {author} {\bibfnamefont {A.}~\bibnamefont {Liang}}, \bibinfo {author}
  {\bibfnamefont {K.}~\bibnamefont {Huang}}, \emph {et~al.},\ }\bibfield
  {title} {\enquote {\bibinfo {title} {Visualization of chiral electronic
  structure and anomalous optical response in a material with chiral charge
  density waves},}\ }\href@noop {} {\bibfield  {journal} {\bibinfo  {journal}
  {Physical review letters}\ }\textbf {\bibinfo {volume} {129}},\ \bibinfo
  {pages} {156401} (\bibinfo {year} {2022})}\BibitemShut {NoStop}%
\bibitem [{\citenamefont {Wang}, \citenamefont {Nepal},\ and\ \citenamefont
  {Canfield}(2024)}]{wang2024}%
  \BibitemOpen
  \bibfield  {author} {\bibinfo {author} {\bibfnamefont {L.-L.}\ \bibnamefont
  {Wang}}, \bibinfo {author} {\bibfnamefont {N.~K.}\ \bibnamefont {Nepal}},\
  and\ \bibinfo {author} {\bibfnamefont {P.~C.}\ \bibnamefont {Canfield}},\
  }\bibfield  {title} {\enquote {\bibinfo {title} {{Origin of charge density
  wave in topological semimetals SrAl4 and EuAl4}},}\ }\href@noop {} {\bibfield
   {journal} {\bibinfo  {journal} {Communications Physics}\ }\textbf {\bibinfo
  {volume} {7}},\ \bibinfo {pages} {111} (\bibinfo {year} {2024})}\BibitemShut
  {NoStop}%
\bibitem [{\citenamefont {Yusupov}\ \emph {et~al.}(2008)\citenamefont
  {Yusupov}, \citenamefont {Mertelj}, \citenamefont {Chu}, \citenamefont
  {Fisher},\ and\ \citenamefont {Mihailovic}}]{yusupov2008}%
  \BibitemOpen
  \bibfield  {author} {\bibinfo {author} {\bibfnamefont {R.}~\bibnamefont
  {Yusupov}}, \bibinfo {author} {\bibfnamefont {T.}~\bibnamefont {Mertelj}},
  \bibinfo {author} {\bibfnamefont {J.-H.}\ \bibnamefont {Chu}}, \bibinfo
  {author} {\bibfnamefont {I.}~\bibnamefont {Fisher}},\ and\ \bibinfo {author}
  {\bibfnamefont {D.}~\bibnamefont {Mihailovic}},\ }\bibfield  {title}
  {\enquote {\bibinfo {title} {{Single-particle and collective mode couplings
  associated with 1-and 2-directional electronic ordering in metallic R Te 3
  (R= Ho, Dy, Tb)}},}\ }\href@noop {} {\bibfield  {journal} {\bibinfo
  {journal} {Physical review letters}\ }\textbf {\bibinfo {volume} {101}},\
  \bibinfo {pages} {246402} (\bibinfo {year} {2008})}\BibitemShut {NoStop}%
\bibitem [{\citenamefont {Schmitt}\ \emph
  {et~al.}(2011{\natexlab{a}})\citenamefont {Schmitt}, \citenamefont
  {Kirchmann}, \citenamefont {Bovensiepen}, \citenamefont {Moore},
  \citenamefont {Chu}, \citenamefont {Lu}, \citenamefont {Rettig},
  \citenamefont {Wolf}, \citenamefont {Fisher},\ and\ \citenamefont
  {Shen}}]{schmitt2011}%
  \BibitemOpen
  \bibfield  {author} {\bibinfo {author} {\bibfnamefont {F.}~\bibnamefont
  {Schmitt}}, \bibinfo {author} {\bibfnamefont {P.~S.}\ \bibnamefont
  {Kirchmann}}, \bibinfo {author} {\bibfnamefont {U.}~\bibnamefont
  {Bovensiepen}}, \bibinfo {author} {\bibfnamefont {R.}~\bibnamefont {Moore}},
  \bibinfo {author} {\bibfnamefont {J.}~\bibnamefont {Chu}}, \bibinfo {author}
  {\bibfnamefont {D.}~\bibnamefont {Lu}}, \bibinfo {author} {\bibfnamefont
  {L.}~\bibnamefont {Rettig}}, \bibinfo {author} {\bibfnamefont
  {M.}~\bibnamefont {Wolf}}, \bibinfo {author} {\bibfnamefont {I.}~\bibnamefont
  {Fisher}},\ and\ \bibinfo {author} {\bibfnamefont {Z.}~\bibnamefont {Shen}},\
  }\bibfield  {title} {\enquote {\bibinfo {title} {{Ultrafast electron dynamics
  in the charge density wave material TbTe3}},}\ }\href@noop {} {\bibfield
  {journal} {\bibinfo  {journal} {New Journal of Physics}\ }\textbf {\bibinfo
  {volume} {13}},\ \bibinfo {pages} {063022} (\bibinfo {year}
  {2011}{\natexlab{a}})}\BibitemShut {NoStop}%
\bibitem [{\citenamefont {Sch{\"a}fer}\ \emph {et~al.}(2010)\citenamefont
  {Sch{\"a}fer}, \citenamefont {Kabanov}, \citenamefont {Beyer}, \citenamefont
  {Biljakovic},\ and\ \citenamefont {Demsar}}]{schaefer2010}%
  \BibitemOpen
  \bibfield  {author} {\bibinfo {author} {\bibfnamefont {H.}~\bibnamefont
  {Sch{\"a}fer}}, \bibinfo {author} {\bibfnamefont {V.~V.}\ \bibnamefont
  {Kabanov}}, \bibinfo {author} {\bibfnamefont {M.}~\bibnamefont {Beyer}},
  \bibinfo {author} {\bibfnamefont {K.}~\bibnamefont {Biljakovic}},\ and\
  \bibinfo {author} {\bibfnamefont {J.}~\bibnamefont {Demsar}},\ }\bibfield
  {title} {\enquote {\bibinfo {title} {Disentanglement of the electronic and
  lattice parts of the order parameter in a 1d charge density wave system
  probed by femtosecond spectroscopy},}\ }\href@noop {} {\bibfield  {journal}
  {\bibinfo  {journal} {Physical review letters}\ }\textbf {\bibinfo {volume}
  {105}},\ \bibinfo {pages} {066402} (\bibinfo {year} {2010})}\BibitemShut
  {NoStop}%
\bibitem [{\citenamefont {Warawa}\ \emph {et~al.}(2023)\citenamefont {Warawa},
  \citenamefont {Christophel}, \citenamefont {Sobolev}, \citenamefont {Demsar},
  \citenamefont {Roskos},\ and\ \citenamefont {Thomson}}]{warawa2023}%
  \BibitemOpen
  \bibfield  {author} {\bibinfo {author} {\bibfnamefont {K.}~\bibnamefont
  {Warawa}}, \bibinfo {author} {\bibfnamefont {N.}~\bibnamefont {Christophel}},
  \bibinfo {author} {\bibfnamefont {S.}~\bibnamefont {Sobolev}}, \bibinfo
  {author} {\bibfnamefont {J.}~\bibnamefont {Demsar}}, \bibinfo {author}
  {\bibfnamefont {H.~G.}\ \bibnamefont {Roskos}},\ and\ \bibinfo {author}
  {\bibfnamefont {M.~D.}\ \bibnamefont {Thomson}},\ }\bibfield  {title}
  {\enquote {\bibinfo {title} {Combined investigation of collective amplitude
  and phase modes in a quasi-one-dimensional charge density wave system over a
  wide spectral range},}\ }\href@noop {} {\bibfield  {journal} {\bibinfo
  {journal} {Physical Review B}\ }\textbf {\bibinfo {volume} {108}},\ \bibinfo
  {pages} {045147} (\bibinfo {year} {2023})}\BibitemShut {NoStop}%
\bibitem [{\citenamefont {Rohwer}\ \emph {et~al.}(2011)\citenamefont {Rohwer},
  \citenamefont {Hellmann}, \citenamefont {Wiesenmayer}, \citenamefont {Sohrt},
  \citenamefont {Stange}, \citenamefont {Slomski}, \citenamefont {Carr},
  \citenamefont {Liu}, \citenamefont {Avila}, \citenamefont {Kall{\"a}ne} \emph
  {et~al.}}]{rohwer2011}%
  \BibitemOpen
  \bibfield  {author} {\bibinfo {author} {\bibfnamefont {T.}~\bibnamefont
  {Rohwer}}, \bibinfo {author} {\bibfnamefont {S.}~\bibnamefont {Hellmann}},
  \bibinfo {author} {\bibfnamefont {M.}~\bibnamefont {Wiesenmayer}}, \bibinfo
  {author} {\bibfnamefont {C.}~\bibnamefont {Sohrt}}, \bibinfo {author}
  {\bibfnamefont {A.}~\bibnamefont {Stange}}, \bibinfo {author} {\bibfnamefont
  {B.}~\bibnamefont {Slomski}}, \bibinfo {author} {\bibfnamefont
  {A.}~\bibnamefont {Carr}}, \bibinfo {author} {\bibfnamefont {Y.}~\bibnamefont
  {Liu}}, \bibinfo {author} {\bibfnamefont {L.~M.}\ \bibnamefont {Avila}},
  \bibinfo {author} {\bibfnamefont {M.}~\bibnamefont {Kall{\"a}ne}}, \emph
  {et~al.},\ }\bibfield  {title} {\enquote {\bibinfo {title} {Collapse of
  long-range charge order tracked by time-resolved photoemission at high
  momenta},}\ }\href@noop {} {\bibfield  {journal} {\bibinfo  {journal}
  {Nature}\ }\textbf {\bibinfo {volume} {471}},\ \bibinfo {pages} {490--493}
  (\bibinfo {year} {2011})}\BibitemShut {NoStop}%
\bibitem [{\citenamefont {Porer}\ \emph {et~al.}(2014)\citenamefont {Porer},
  \citenamefont {Leierseder}, \citenamefont {M{\'e}nard}, \citenamefont
  {Dachraoui}, \citenamefont {Mouchliadis}, \citenamefont {Perakis},
  \citenamefont {Heinzmann}, \citenamefont {Demsar}, \citenamefont
  {Rossnagel},\ and\ \citenamefont {Huber}}]{porer2014}%
  \BibitemOpen
  \bibfield  {author} {\bibinfo {author} {\bibfnamefont {M.}~\bibnamefont
  {Porer}}, \bibinfo {author} {\bibfnamefont {U.}~\bibnamefont {Leierseder}},
  \bibinfo {author} {\bibfnamefont {J.-M.}\ \bibnamefont {M{\'e}nard}},
  \bibinfo {author} {\bibfnamefont {H.}~\bibnamefont {Dachraoui}}, \bibinfo
  {author} {\bibfnamefont {L.}~\bibnamefont {Mouchliadis}}, \bibinfo {author}
  {\bibfnamefont {I.}~\bibnamefont {Perakis}}, \bibinfo {author} {\bibfnamefont
  {U.}~\bibnamefont {Heinzmann}}, \bibinfo {author} {\bibfnamefont
  {J.}~\bibnamefont {Demsar}}, \bibinfo {author} {\bibfnamefont
  {K.}~\bibnamefont {Rossnagel}},\ and\ \bibinfo {author} {\bibfnamefont
  {R.}~\bibnamefont {Huber}},\ }\bibfield  {title} {\enquote {\bibinfo {title}
  {Non-thermal separation of electronic and structural orders in a persisting
  charge density wave},}\ }\href@noop {} {\bibfield  {journal} {\bibinfo
  {journal} {Nature materials}\ }\textbf {\bibinfo {volume} {13}},\ \bibinfo
  {pages} {857--861} (\bibinfo {year} {2014})}\BibitemShut {NoStop}%
\bibitem [{\citenamefont {Frigge}\ \emph {et~al.}(2017)\citenamefont {Frigge},
  \citenamefont {Hafke}, \citenamefont {Witte}, \citenamefont {Krenzer},
  \citenamefont {Streub{\"u}hr}, \citenamefont {Samad~Syed}, \citenamefont
  {Mik{\v{s}}i{\'c}~Trontl}, \citenamefont {Avigo}, \citenamefont {Zhou},
  \citenamefont {Ligges} \emph {et~al.}}]{frigge2017}%
  \BibitemOpen
  \bibfield  {author} {\bibinfo {author} {\bibfnamefont {T.}~\bibnamefont
  {Frigge}}, \bibinfo {author} {\bibfnamefont {B.}~\bibnamefont {Hafke}},
  \bibinfo {author} {\bibfnamefont {T.}~\bibnamefont {Witte}}, \bibinfo
  {author} {\bibfnamefont {B.}~\bibnamefont {Krenzer}}, \bibinfo {author}
  {\bibfnamefont {C.}~\bibnamefont {Streub{\"u}hr}}, \bibinfo {author}
  {\bibfnamefont {A.}~\bibnamefont {Samad~Syed}}, \bibinfo {author}
  {\bibfnamefont {V.}~\bibnamefont {Mik{\v{s}}i{\'c}~Trontl}}, \bibinfo
  {author} {\bibfnamefont {I.}~\bibnamefont {Avigo}}, \bibinfo {author}
  {\bibfnamefont {P.}~\bibnamefont {Zhou}}, \bibinfo {author} {\bibfnamefont
  {M.}~\bibnamefont {Ligges}}, \emph {et~al.},\ }\bibfield  {title} {\enquote
  {\bibinfo {title} {Optically excited structural transition in atomic wires on
  surfaces at the quantum limit},}\ }\href@noop {} {\bibfield  {journal}
  {\bibinfo  {journal} {Nature}\ }\textbf {\bibinfo {volume} {544}},\ \bibinfo
  {pages} {207--211} (\bibinfo {year} {2017})}\BibitemShut {NoStop}%
\bibitem [{\citenamefont {Eichberger}\ \emph {et~al.}(2010)\citenamefont
  {Eichberger}, \citenamefont {Sch{\"a}fer}, \citenamefont {Krumova},
  \citenamefont {Beyer}, \citenamefont {Demsar}, \citenamefont {Berger},
  \citenamefont {Moriena}, \citenamefont {Sciaini},\ and\ \citenamefont
  {Miller}}]{eichberger2010}%
  \BibitemOpen
  \bibfield  {author} {\bibinfo {author} {\bibfnamefont {M.}~\bibnamefont
  {Eichberger}}, \bibinfo {author} {\bibfnamefont {H.}~\bibnamefont
  {Sch{\"a}fer}}, \bibinfo {author} {\bibfnamefont {M.}~\bibnamefont
  {Krumova}}, \bibinfo {author} {\bibfnamefont {M.}~\bibnamefont {Beyer}},
  \bibinfo {author} {\bibfnamefont {J.}~\bibnamefont {Demsar}}, \bibinfo
  {author} {\bibfnamefont {H.}~\bibnamefont {Berger}}, \bibinfo {author}
  {\bibfnamefont {G.}~\bibnamefont {Moriena}}, \bibinfo {author} {\bibfnamefont
  {G.}~\bibnamefont {Sciaini}},\ and\ \bibinfo {author} {\bibfnamefont {R.~D.}\
  \bibnamefont {Miller}},\ }\bibfield  {title} {\enquote {\bibinfo {title}
  {Snapshots of cooperative atomic motions in the optical suppression of charge
  density waves},}\ }\href@noop {} {\bibfield  {journal} {\bibinfo  {journal}
  {Nature}\ }\textbf {\bibinfo {volume} {468}},\ \bibinfo {pages} {799--802}
  (\bibinfo {year} {2010})}\BibitemShut {NoStop}%
\bibitem [{\citenamefont {Haupt}\ \emph {et~al.}(2016)\citenamefont {Haupt},
  \citenamefont {Eichberger}, \citenamefont {Erasmus}, \citenamefont {Rohwer},
  \citenamefont {Demsar}, \citenamefont {Rossnagel},\ and\ \citenamefont
  {Schwoerer}}]{haupt2016}%
  \BibitemOpen
  \bibfield  {author} {\bibinfo {author} {\bibfnamefont {K.}~\bibnamefont
  {Haupt}}, \bibinfo {author} {\bibfnamefont {M.}~\bibnamefont {Eichberger}},
  \bibinfo {author} {\bibfnamefont {N.}~\bibnamefont {Erasmus}}, \bibinfo
  {author} {\bibfnamefont {A.}~\bibnamefont {Rohwer}}, \bibinfo {author}
  {\bibfnamefont {J.}~\bibnamefont {Demsar}}, \bibinfo {author} {\bibfnamefont
  {K.}~\bibnamefont {Rossnagel}},\ and\ \bibinfo {author} {\bibfnamefont
  {H.}~\bibnamefont {Schwoerer}},\ }\bibfield  {title} {\enquote {\bibinfo
  {title} {Ultrafast metamorphosis of a complex charge-density wave},}\
  }\href@noop {} {\bibfield  {journal} {\bibinfo  {journal} {Physical review
  letters}\ }\textbf {\bibinfo {volume} {116}},\ \bibinfo {pages} {016402}
  (\bibinfo {year} {2016})}\BibitemShut {NoStop}%
\bibitem [{\citenamefont {Danz}, \citenamefont {Domr{\"o}se},\ and\
  \citenamefont {Ropers}(2021)}]{danz2021}%
  \BibitemOpen
  \bibfield  {author} {\bibinfo {author} {\bibfnamefont {T.}~\bibnamefont
  {Danz}}, \bibinfo {author} {\bibfnamefont {T.}~\bibnamefont {Domr{\"o}se}},\
  and\ \bibinfo {author} {\bibfnamefont {C.}~\bibnamefont {Ropers}},\
  }\bibfield  {title} {\enquote {\bibinfo {title} {Ultrafast nanoimaging of the
  order parameter in a structural phase transition},}\ }\href@noop {}
  {\bibfield  {journal} {\bibinfo  {journal} {Science}\ }\textbf {\bibinfo
  {volume} {371}},\ \bibinfo {pages} {371--374} (\bibinfo {year}
  {2021})}\BibitemShut {NoStop}%
\bibitem [{\citenamefont {Stojchevska}\ \emph {et~al.}(2014)\citenamefont
  {Stojchevska}, \citenamefont {Vaskivskyi}, \citenamefont {Mertelj},
  \citenamefont {Kusar}, \citenamefont {Svetin}, \citenamefont {Brazovskii},\
  and\ \citenamefont {Mihailovic}}]{stojchevska2014}%
  \BibitemOpen
  \bibfield  {author} {\bibinfo {author} {\bibfnamefont {L.}~\bibnamefont
  {Stojchevska}}, \bibinfo {author} {\bibfnamefont {I.}~\bibnamefont
  {Vaskivskyi}}, \bibinfo {author} {\bibfnamefont {T.}~\bibnamefont {Mertelj}},
  \bibinfo {author} {\bibfnamefont {P.}~\bibnamefont {Kusar}}, \bibinfo
  {author} {\bibfnamefont {D.}~\bibnamefont {Svetin}}, \bibinfo {author}
  {\bibfnamefont {S.}~\bibnamefont {Brazovskii}},\ and\ \bibinfo {author}
  {\bibfnamefont {D.}~\bibnamefont {Mihailovic}},\ }\bibfield  {title}
  {\enquote {\bibinfo {title} {Ultrafast switching to a stable hidden quantum
  state in an electronic crystal},}\ }\href@noop {} {\bibfield  {journal}
  {\bibinfo  {journal} {Science}\ }\textbf {\bibinfo {volume} {344}},\ \bibinfo
  {pages} {177--180} (\bibinfo {year} {2014})}\BibitemShut {NoStop}%
\bibitem [{\citenamefont {Vaskivskyi}\ \emph {et~al.}(2015)\citenamefont
  {Vaskivskyi}, \citenamefont {Gospodaric}, \citenamefont {Brazovskii},
  \citenamefont {Svetin}, \citenamefont {Sutar}, \citenamefont {Goreshnik},
  \citenamefont {Mihailovic}, \citenamefont {Mertelj},\ and\ \citenamefont
  {Mihailovic}}]{vaskivskyi2015}%
  \BibitemOpen
  \bibfield  {author} {\bibinfo {author} {\bibfnamefont {I.}~\bibnamefont
  {Vaskivskyi}}, \bibinfo {author} {\bibfnamefont {J.}~\bibnamefont
  {Gospodaric}}, \bibinfo {author} {\bibfnamefont {S.}~\bibnamefont
  {Brazovskii}}, \bibinfo {author} {\bibfnamefont {D.}~\bibnamefont {Svetin}},
  \bibinfo {author} {\bibfnamefont {P.}~\bibnamefont {Sutar}}, \bibinfo
  {author} {\bibfnamefont {E.}~\bibnamefont {Goreshnik}}, \bibinfo {author}
  {\bibfnamefont {I.~A.}\ \bibnamefont {Mihailovic}}, \bibinfo {author}
  {\bibfnamefont {T.}~\bibnamefont {Mertelj}},\ and\ \bibinfo {author}
  {\bibfnamefont {D.}~\bibnamefont {Mihailovic}},\ }\bibfield  {title}
  {\enquote {\bibinfo {title} {{Controlling the metal-to-insulator relaxation
  of the metastable hidden quantum state in 1T-TaS2}},}\ }\href@noop {}
  {\bibfield  {journal} {\bibinfo  {journal} {Science advances}\ }\textbf
  {\bibinfo {volume} {1}},\ \bibinfo {pages} {e1500168} (\bibinfo {year}
  {2015})}\BibitemShut {NoStop}%
\bibitem [{\citenamefont {Maklar}\ \emph {et~al.}(2023)\citenamefont {Maklar},
  \citenamefont {Sarkar}, \citenamefont {Dong}, \citenamefont {Gerasimenko},
  \citenamefont {Pincelli}, \citenamefont {Beaulieu}, \citenamefont
  {Kirchmann}, \citenamefont {Sobota}, \citenamefont {Yang}, \citenamefont
  {Leuenberger} \emph {et~al.}}]{maklar2023}%
  \BibitemOpen
  \bibfield  {author} {\bibinfo {author} {\bibfnamefont {J.}~\bibnamefont
  {Maklar}}, \bibinfo {author} {\bibfnamefont {J.}~\bibnamefont {Sarkar}},
  \bibinfo {author} {\bibfnamefont {S.}~\bibnamefont {Dong}}, \bibinfo {author}
  {\bibfnamefont {Y.~A.}\ \bibnamefont {Gerasimenko}}, \bibinfo {author}
  {\bibfnamefont {T.}~\bibnamefont {Pincelli}}, \bibinfo {author}
  {\bibfnamefont {S.}~\bibnamefont {Beaulieu}}, \bibinfo {author}
  {\bibfnamefont {P.~S.}\ \bibnamefont {Kirchmann}}, \bibinfo {author}
  {\bibfnamefont {J.~A.}\ \bibnamefont {Sobota}}, \bibinfo {author}
  {\bibfnamefont {S.}~\bibnamefont {Yang}}, \bibinfo {author} {\bibfnamefont
  {D.}~\bibnamefont {Leuenberger}}, \emph {et~al.},\ }\bibfield  {title}
  {\enquote {\bibinfo {title} {Coherent light control of a metastable hidden
  state},}\ }\href@noop {} {\bibfield  {journal} {\bibinfo  {journal} {Science
  advances}\ }\textbf {\bibinfo {volume} {9}},\ \bibinfo {pages} {eadi4661}
  (\bibinfo {year} {2023})}\BibitemShut {NoStop}%
\bibitem [{\citenamefont {Sacchetti}\ \emph {et~al.}(2006)\citenamefont
  {Sacchetti}, \citenamefont {Degiorgi}, \citenamefont {Giamarchi},
  \citenamefont {Ru},\ and\ \citenamefont {Fisher}}]{sacchetti2006}%
  \BibitemOpen
  \bibfield  {author} {\bibinfo {author} {\bibfnamefont {A.}~\bibnamefont
  {Sacchetti}}, \bibinfo {author} {\bibfnamefont {L.}~\bibnamefont {Degiorgi}},
  \bibinfo {author} {\bibfnamefont {T.}~\bibnamefont {Giamarchi}}, \bibinfo
  {author} {\bibfnamefont {N.}~\bibnamefont {Ru}},\ and\ \bibinfo {author}
  {\bibfnamefont {I.}~\bibnamefont {Fisher}},\ }\bibfield  {title} {\enquote
  {\bibinfo {title} {Chemical pressure and hidden one-dimensional behavior in
  rare-earth tri-telluride charge-density wave compounds},}\ }\href@noop {}
  {\bibfield  {journal} {\bibinfo  {journal} {Physical Review B—Condensed
  Matter and Materials Physics}\ }\textbf {\bibinfo {volume} {74}},\ \bibinfo
  {pages} {125115} (\bibinfo {year} {2006})}\BibitemShut {NoStop}%
\bibitem [{\citenamefont {Hu}\ \emph {et~al.}(2014)\citenamefont {Hu},
  \citenamefont {Cheng}, \citenamefont {Yuan}, \citenamefont {Dong},\ and\
  \citenamefont {Wang}}]{NLWang2014}%
  \BibitemOpen
  \bibfield  {author} {\bibinfo {author} {\bibfnamefont {B.~F.}\ \bibnamefont
  {Hu}}, \bibinfo {author} {\bibfnamefont {B.}~\bibnamefont {Cheng}}, \bibinfo
  {author} {\bibfnamefont {R.~H.}\ \bibnamefont {Yuan}}, \bibinfo {author}
  {\bibfnamefont {T.}~\bibnamefont {Dong}},\ and\ \bibinfo {author}
  {\bibfnamefont {N.~L.}\ \bibnamefont {Wang}},\ }\bibfield  {title} {\enquote
  {\bibinfo {title} {Coexistence and competition of multiple
  charge-density-wave orders in rare-earth tritellurides},}\ }\href
  {https://doi.org/10.1103/PhysRevB.90.085105} {\bibfield  {journal} {\bibinfo
  {journal} {Phys. Rev. B}\ }\textbf {\bibinfo {volume} {90}},\ \bibinfo
  {pages} {085105} (\bibinfo {year} {2014})}\BibitemShut {NoStop}%
\bibitem [{\citenamefont {Tuniz}\ \emph {et~al.}(2025)\citenamefont {Tuniz},
  \citenamefont {Consiglio}, \citenamefont {Pokharel}, \citenamefont
  {Parmigiani}, \citenamefont {Neupert}, \citenamefont {Thomale}, \citenamefont
  {Chaluvadi}, \citenamefont {Orgiani}, \citenamefont {Sangiovanni},
  \citenamefont {Wilson} \emph {et~al.}}]{tuniz2025}%
  \BibitemOpen
  \bibfield  {author} {\bibinfo {author} {\bibfnamefont {M.}~\bibnamefont
  {Tuniz}}, \bibinfo {author} {\bibfnamefont {A.}~\bibnamefont {Consiglio}},
  \bibinfo {author} {\bibfnamefont {G.}~\bibnamefont {Pokharel}}, \bibinfo
  {author} {\bibfnamefont {F.}~\bibnamefont {Parmigiani}}, \bibinfo {author}
  {\bibfnamefont {T.}~\bibnamefont {Neupert}}, \bibinfo {author} {\bibfnamefont
  {R.}~\bibnamefont {Thomale}}, \bibinfo {author} {\bibfnamefont {S.~K.}\
  \bibnamefont {Chaluvadi}}, \bibinfo {author} {\bibfnamefont {P.}~\bibnamefont
  {Orgiani}}, \bibinfo {author} {\bibfnamefont {G.}~\bibnamefont
  {Sangiovanni}}, \bibinfo {author} {\bibfnamefont {S.~D.}\ \bibnamefont
  {Wilson}}, \emph {et~al.},\ }\bibfield  {title} {\enquote {\bibinfo {title}
  {{Strain-Induced Enhancement of the Charge Density Wave in the Kagome Metal
  ScV6Sn6}},}\ }\href@noop {} {\bibfield  {journal} {\bibinfo  {journal}
  {Physical Review Letters}\ }\textbf {\bibinfo {volume} {134}},\ \bibinfo
  {pages} {066501} (\bibinfo {year} {2025})}\BibitemShut {NoStop}%
\bibitem [{\citenamefont {Lanz}\ \emph {et~al.}()\citenamefont {Lanz},
  \citenamefont {Tkach}, \citenamefont {Lytvynenko}, \citenamefont {Diekmann},
  \citenamefont {Agarwal}, \citenamefont {Souliou}, \citenamefont {Frachet},
  \citenamefont {Merz}, \citenamefont {Chernov}, \citenamefont {Gloskovskii},
  \citenamefont {Koller}, \citenamefont {Kutnyakhov}, \citenamefont {Hoesch},
  \citenamefont {Schlueter}, \citenamefont {Rüssmann}, \citenamefont
  {Mokrousov}, \citenamefont {Le~Tacon}, \citenamefont {Rossnagel},
  \citenamefont {Demsar}, \citenamefont {Schönhense}, \citenamefont {Elmers},\
  and\ \citenamefont {Fedchenko}}]{lanz2026}%
  \BibitemOpen
  \bibfield  {author} {\bibinfo {author} {\bibfnamefont {A.}~\bibnamefont
  {Lanz}}, \bibinfo {author} {\bibfnamefont {O.}~\bibnamefont {Tkach}},
  \bibinfo {author} {\bibfnamefont {Y.}~\bibnamefont {Lytvynenko}}, \bibinfo
  {author} {\bibfnamefont {F.}~\bibnamefont {Diekmann}}, \bibinfo {author}
  {\bibfnamefont {H.}~\bibnamefont {Agarwal}}, \bibinfo {author} {\bibfnamefont
  {S.~M.}\ \bibnamefont {Souliou}}, \bibinfo {author} {\bibfnamefont
  {M.}~\bibnamefont {Frachet}}, \bibinfo {author} {\bibfnamefont
  {M.}~\bibnamefont {Merz}}, \bibinfo {author} {\bibfnamefont {S.~V.}\
  \bibnamefont {Chernov}}, \bibinfo {author} {\bibfnamefont {A.}~\bibnamefont
  {Gloskovskii}}, \bibinfo {author} {\bibfnamefont {V.}~\bibnamefont {Koller}},
  \bibinfo {author} {\bibfnamefont {D.}~\bibnamefont {Kutnyakhov}}, \bibinfo
  {author} {\bibfnamefont {M.}~\bibnamefont {Hoesch}}, \bibinfo {author}
  {\bibfnamefont {C.}~\bibnamefont {Schlueter}}, \bibinfo {author}
  {\bibfnamefont {P.}~\bibnamefont {Rüssmann}}, \bibinfo {author}
  {\bibfnamefont {Y.}~\bibnamefont {Mokrousov}}, \bibinfo {author}
  {\bibfnamefont {M.}~\bibnamefont {Le~Tacon}}, \bibinfo {author}
  {\bibfnamefont {K.}~\bibnamefont {Rossnagel}}, \bibinfo {author}
  {\bibfnamefont {J.}~\bibnamefont {Demsar}}, \bibinfo {author} {\bibfnamefont
  {G.}~\bibnamefont {Schönhense}}, \bibinfo {author} {\bibfnamefont {H.-J.}\
  \bibnamefont {Elmers}},\ and\ \bibinfo {author} {\bibfnamefont
  {O.}~\bibnamefont {Fedchenko}},\ }\bibfield  {title} {\enquote {\bibinfo
  {title} {Strain tuning the occupation of candidate topological weyl states in
  w-doped mote2},}\ }\href {https://doi.org/https://doi.org/10.1002/advs.76064}
  {\bibfield  {journal} {\bibinfo  {journal} {Advanced Science}\ }\textbf
  {\bibinfo {volume} {n/a}},\ \bibinfo {pages} {e76064}},\ \Eprint
  {https://arxiv.org/abs/https://advanced.onlinelibrary.wiley.com/doi/pdf/10.1002/advs.76064}
  {https://advanced.onlinelibrary.wiley.com/doi/pdf/10.1002/advs.76064}
  \BibitemShut {NoStop}%
\bibitem [{\citenamefont {Sacchetti}\ \emph {et~al.}(2007)\citenamefont
  {Sacchetti}, \citenamefont {Arcangeletti}, \citenamefont {Perucchi},
  \citenamefont {Baldassarre}, \citenamefont {Postorino}, \citenamefont {Lupi},
  \citenamefont {Ru}, \citenamefont {Fisher},\ and\ \citenamefont
  {Degiorgi}}]{sacchetti2007}%
  \BibitemOpen
  \bibfield  {author} {\bibinfo {author} {\bibfnamefont {A.}~\bibnamefont
  {Sacchetti}}, \bibinfo {author} {\bibfnamefont {E.}~\bibnamefont
  {Arcangeletti}}, \bibinfo {author} {\bibfnamefont {A.}~\bibnamefont
  {Perucchi}}, \bibinfo {author} {\bibfnamefont {L.}~\bibnamefont
  {Baldassarre}}, \bibinfo {author} {\bibfnamefont {P.}~\bibnamefont
  {Postorino}}, \bibinfo {author} {\bibfnamefont {S.}~\bibnamefont {Lupi}},
  \bibinfo {author} {\bibfnamefont {N.}~\bibnamefont {Ru}}, \bibinfo {author}
  {\bibfnamefont {I.}~\bibnamefont {Fisher}},\ and\ \bibinfo {author}
  {\bibfnamefont {L.}~\bibnamefont {Degiorgi}},\ }\bibfield  {title} {\enquote
  {\bibinfo {title} {Pressure dependence of the charge-density-wave gap in
  rare-earth tritellurides},}\ }\href@noop {} {\bibfield  {journal} {\bibinfo
  {journal} {Physical review letters}\ }\textbf {\bibinfo {volume} {98}},\
  \bibinfo {pages} {026401} (\bibinfo {year} {2007})}\BibitemShut {NoStop}%
\bibitem [{\citenamefont {Walmsley}\ \emph {et~al.}(2020)\citenamefont
  {Walmsley}, \citenamefont {Aeschlimann}, \citenamefont {Straquadine},
  \citenamefont {Giraldo-Gallo}, \citenamefont {Riggs}, \citenamefont {Chan},
  \citenamefont {McDonald},\ and\ \citenamefont {Fisher}}]{walmsley2020}%
  \BibitemOpen
  \bibfield  {author} {\bibinfo {author} {\bibfnamefont {P.}~\bibnamefont
  {Walmsley}}, \bibinfo {author} {\bibfnamefont {S.}~\bibnamefont
  {Aeschlimann}}, \bibinfo {author} {\bibfnamefont {J.~A.~W.}\ \bibnamefont
  {Straquadine}}, \bibinfo {author} {\bibfnamefont {P.}~\bibnamefont
  {Giraldo-Gallo}}, \bibinfo {author} {\bibfnamefont {S.~C.}\ \bibnamefont
  {Riggs}}, \bibinfo {author} {\bibfnamefont {M.~K.}\ \bibnamefont {Chan}},
  \bibinfo {author} {\bibfnamefont {R.~D.}\ \bibnamefont {McDonald}},\ and\
  \bibinfo {author} {\bibfnamefont {I.~R.}\ \bibnamefont {Fisher}},\ }\bibfield
   {title} {\enquote {\bibinfo {title} {Magnetic breakdown and charge density
  wave formation: A quantum oscillation study of the rare-earth
  tritellurides},}\ }\href {https://doi.org/10.1103/PhysRevB.102.045150}
  {\bibfield  {journal} {\bibinfo  {journal} {Phys. Rev. B}\ }\textbf {\bibinfo
  {volume} {102}},\ \bibinfo {pages} {045150} (\bibinfo {year}
  {2020})}\BibitemShut {NoStop}%
\bibitem [{\citenamefont {Yumigeta}\ \emph {et~al.}(2021)\citenamefont
  {Yumigeta}, \citenamefont {Qin}, \citenamefont {Li}, \citenamefont {Blei},
  \citenamefont {Attarde}, \citenamefont {Kopas},\ and\ \citenamefont
  {Tongay}}]{yumigeta2021}%
  \BibitemOpen
  \bibfield  {author} {\bibinfo {author} {\bibfnamefont {K.}~\bibnamefont
  {Yumigeta}}, \bibinfo {author} {\bibfnamefont {Y.}~\bibnamefont {Qin}},
  \bibinfo {author} {\bibfnamefont {H.}~\bibnamefont {Li}}, \bibinfo {author}
  {\bibfnamefont {M.}~\bibnamefont {Blei}}, \bibinfo {author} {\bibfnamefont
  {Y.}~\bibnamefont {Attarde}}, \bibinfo {author} {\bibfnamefont
  {C.}~\bibnamefont {Kopas}},\ and\ \bibinfo {author} {\bibfnamefont
  {S.}~\bibnamefont {Tongay}},\ }\bibfield  {title} {\enquote {\bibinfo {title}
  {Advances in rare-earth tritelluride quantum materials: Structure,
  properties, and synthesis},}\ }\href@noop {} {\bibfield  {journal} {\bibinfo
  {journal} {Advanced Science}\ }\textbf {\bibinfo {volume} {8}},\ \bibinfo
  {pages} {2004762} (\bibinfo {year} {2021})}\BibitemShut {NoStop}%
\bibitem [{\citenamefont {Brouet}\ \emph {et~al.}(2008)\citenamefont {Brouet},
  \citenamefont {Yang}, \citenamefont {Zhou}, \citenamefont {Hussain},
  \citenamefont {Moore}, \citenamefont {He}, \citenamefont {Lu}, \citenamefont
  {Shen}, \citenamefont {Laverock}, \citenamefont {Dugdale}, \citenamefont
  {Ru},\ and\ \citenamefont {Fisher}}]{ARPES2008}%
  \BibitemOpen
  \bibfield  {author} {\bibinfo {author} {\bibfnamefont {V.}~\bibnamefont
  {Brouet}}, \bibinfo {author} {\bibfnamefont {W.~L.}\ \bibnamefont {Yang}},
  \bibinfo {author} {\bibfnamefont {X.~J.}\ \bibnamefont {Zhou}}, \bibinfo
  {author} {\bibfnamefont {Z.}~\bibnamefont {Hussain}}, \bibinfo {author}
  {\bibfnamefont {R.~G.}\ \bibnamefont {Moore}}, \bibinfo {author}
  {\bibfnamefont {R.}~\bibnamefont {He}}, \bibinfo {author} {\bibfnamefont
  {D.~H.}\ \bibnamefont {Lu}}, \bibinfo {author} {\bibfnamefont {Z.~X.}\
  \bibnamefont {Shen}}, \bibinfo {author} {\bibfnamefont {J.}~\bibnamefont
  {Laverock}}, \bibinfo {author} {\bibfnamefont {S.~B.}\ \bibnamefont
  {Dugdale}}, \bibinfo {author} {\bibfnamefont {N.}~\bibnamefont {Ru}},\ and\
  \bibinfo {author} {\bibfnamefont {I.~R.}\ \bibnamefont {Fisher}},\ }\bibfield
   {title} {\enquote {\bibinfo {title} {{Angle-resolved photoemission study of
  the evolution of band structure and charge density wave properties in
  $R{\text{Te}}_{3}$ ($R=\text{Y}$, La, Ce, Sm, Gd, Tb, and Dy)}},}\ }\href
  {https://doi.org/10.1103/PhysRevB.77.235104} {\bibfield  {journal} {\bibinfo
  {journal} {Phys. Rev. B}\ }\textbf {\bibinfo {volume} {77}},\ \bibinfo
  {pages} {235104} (\bibinfo {year} {2008})}\BibitemShut {NoStop}%
\bibitem [{\citenamefont {Eiter}\ \emph {et~al.}(2013)\citenamefont {Eiter},
  \citenamefont {Lavagnini}, \citenamefont {Hackl}, \citenamefont {Nowadnick},
  \citenamefont {Kemper}, \citenamefont {Devereaux}, \citenamefont {Chu},
  \citenamefont {Analytis}, \citenamefont {Fisher},\ and\ \citenamefont
  {Degiorgi}}]{eiter2012}%
  \BibitemOpen
  \bibfield  {author} {\bibinfo {author} {\bibfnamefont {H.-M.}\ \bibnamefont
  {Eiter}}, \bibinfo {author} {\bibfnamefont {M.}~\bibnamefont {Lavagnini}},
  \bibinfo {author} {\bibfnamefont {R.}~\bibnamefont {Hackl}}, \bibinfo
  {author} {\bibfnamefont {E.~A.}\ \bibnamefont {Nowadnick}}, \bibinfo {author}
  {\bibfnamefont {A.~F.}\ \bibnamefont {Kemper}}, \bibinfo {author}
  {\bibfnamefont {T.~P.}\ \bibnamefont {Devereaux}}, \bibinfo {author}
  {\bibfnamefont {J.-H.}\ \bibnamefont {Chu}}, \bibinfo {author} {\bibfnamefont
  {J.~G.}\ \bibnamefont {Analytis}}, \bibinfo {author} {\bibfnamefont {I.~R.}\
  \bibnamefont {Fisher}},\ and\ \bibinfo {author} {\bibfnamefont
  {L.}~\bibnamefont {Degiorgi}},\ }\bibfield  {title} {\enquote {\bibinfo
  {title} {Alternative route to charge density wave formation in multiband
  systems},}\ }\href {https://doi.org/10.1073/pnas.1214745110} {\bibfield
  {journal} {\bibinfo  {journal} {Proceedings of the National Academy of
  Sciences}\ }\textbf {\bibinfo {volume} {110}},\ \bibinfo {pages} {64--69}
  (\bibinfo {year} {2013})},\ \Eprint
  {https://arxiv.org/abs/https://www.pnas.org/doi/pdf/10.1073/pnas.1214745110}
  {https://www.pnas.org/doi/pdf/10.1073/pnas.1214745110} \BibitemShut {NoStop}%
\bibitem [{\citenamefont {Maschek}\ \emph {et~al.}(2015)\citenamefont
  {Maschek}, \citenamefont {Rosenkranz}, \citenamefont {Heid}, \citenamefont
  {Said}, \citenamefont {Giraldo-Gallo}, \citenamefont {Fisher},\ and\
  \citenamefont {Weber}}]{maschek2015}%
  \BibitemOpen
  \bibfield  {author} {\bibinfo {author} {\bibfnamefont {M.}~\bibnamefont
  {Maschek}}, \bibinfo {author} {\bibfnamefont {S.}~\bibnamefont {Rosenkranz}},
  \bibinfo {author} {\bibfnamefont {R.}~\bibnamefont {Heid}}, \bibinfo {author}
  {\bibfnamefont {A.~H.}\ \bibnamefont {Said}}, \bibinfo {author}
  {\bibfnamefont {P.}~\bibnamefont {Giraldo-Gallo}}, \bibinfo {author}
  {\bibfnamefont {I.~R.}\ \bibnamefont {Fisher}},\ and\ \bibinfo {author}
  {\bibfnamefont {F.}~\bibnamefont {Weber}},\ }\bibfield  {title} {\enquote
  {\bibinfo {title} {Wave-vector-dependent electron-phonon coupling and the
  charge-density-wave transition in $\mathrm{TbT}{\mathrm{e}}_{3}$},}\ }\href
  {https://doi.org/10.1103/PhysRevB.91.235146} {\bibfield  {journal} {\bibinfo
  {journal} {Phys. Rev. B}\ }\textbf {\bibinfo {volume} {91}},\ \bibinfo
  {pages} {235146} (\bibinfo {year} {2015})}\BibitemShut {NoStop}%
\bibitem [{\citenamefont {Maschek}\ \emph {et~al.}(2018)\citenamefont
  {Maschek}, \citenamefont {Zocco}, \citenamefont {Rosenkranz}, \citenamefont
  {Heid}, \citenamefont {Said}, \citenamefont {Alatas}, \citenamefont
  {Walmsley}, \citenamefont {Fisher},\ and\ \citenamefont
  {Weber}}]{maschek2018}%
  \BibitemOpen
  \bibfield  {author} {\bibinfo {author} {\bibfnamefont {M.}~\bibnamefont
  {Maschek}}, \bibinfo {author} {\bibfnamefont {D.~A.}\ \bibnamefont {Zocco}},
  \bibinfo {author} {\bibfnamefont {S.}~\bibnamefont {Rosenkranz}}, \bibinfo
  {author} {\bibfnamefont {R.}~\bibnamefont {Heid}}, \bibinfo {author}
  {\bibfnamefont {A.~H.}\ \bibnamefont {Said}}, \bibinfo {author}
  {\bibfnamefont {A.}~\bibnamefont {Alatas}}, \bibinfo {author} {\bibfnamefont
  {P.}~\bibnamefont {Walmsley}}, \bibinfo {author} {\bibfnamefont {I.~R.}\
  \bibnamefont {Fisher}},\ and\ \bibinfo {author} {\bibfnamefont
  {F.}~\bibnamefont {Weber}},\ }\bibfield  {title} {\enquote {\bibinfo {title}
  {Competing soft phonon modes at the charge-density-wave transitions in
  $\mathrm{DyT}{\mathrm{e}}_{3}$},}\ }\href
  {https://doi.org/10.1103/PhysRevB.98.094304} {\bibfield  {journal} {\bibinfo
  {journal} {Phys. Rev. B}\ }\textbf {\bibinfo {volume} {98}},\ \bibinfo
  {pages} {094304} (\bibinfo {year} {2018})}\BibitemShut {NoStop}%
\bibitem [{\citenamefont {Sacchetti}\ \emph {et~al.}(2009)\citenamefont
  {Sacchetti}, \citenamefont {Condron}, \citenamefont {Gvasaliya},
  \citenamefont {Pfuner}, \citenamefont {Lavagnini}, \citenamefont {Baldini},
  \citenamefont {Toney}, \citenamefont {Merlini}, \citenamefont {Hanfland},
  \citenamefont {Mesot} \emph {et~al.}}]{sacchetti2009}%
  \BibitemOpen
  \bibfield  {author} {\bibinfo {author} {\bibfnamefont {A.}~\bibnamefont
  {Sacchetti}}, \bibinfo {author} {\bibfnamefont {C.}~\bibnamefont {Condron}},
  \bibinfo {author} {\bibfnamefont {S.}~\bibnamefont {Gvasaliya}}, \bibinfo
  {author} {\bibfnamefont {F.}~\bibnamefont {Pfuner}}, \bibinfo {author}
  {\bibfnamefont {M.}~\bibnamefont {Lavagnini}}, \bibinfo {author}
  {\bibfnamefont {M.}~\bibnamefont {Baldini}}, \bibinfo {author} {\bibfnamefont
  {M.}~\bibnamefont {Toney}}, \bibinfo {author} {\bibfnamefont
  {M.}~\bibnamefont {Merlini}}, \bibinfo {author} {\bibfnamefont
  {M.}~\bibnamefont {Hanfland}}, \bibinfo {author} {\bibfnamefont
  {J.}~\bibnamefont {Mesot}}, \emph {et~al.},\ }\bibfield  {title} {\enquote
  {\bibinfo {title} {Pressure-induced quenching of the charge-density-wave
  state in rare-earth tritellurides observed by x-ray diffraction},}\
  }\href@noop {} {\bibfield  {journal} {\bibinfo  {journal} {Physical Review
  B—Condensed Matter and Materials Physics}\ }\textbf {\bibinfo {volume}
  {79}},\ \bibinfo {pages} {201101} (\bibinfo {year} {2009})}\BibitemShut
  {NoStop}%
\bibitem [{\citenamefont {Li}\ \emph {et~al.}(2024)\citenamefont {Li},
  \citenamefont {Feng}, \citenamefont {Wang}, \citenamefont {Peng},
  \citenamefont {Li}, \citenamefont {Wang}, \citenamefont {Xu}, \citenamefont
  {Zhao}, \citenamefont {Zhao}, \citenamefont {Jiang} \emph {et~al.}}]{li2024}%
  \BibitemOpen
  \bibfield  {author} {\bibinfo {author} {\bibfnamefont {J.}~\bibnamefont
  {Li}}, \bibinfo {author} {\bibfnamefont {J.}~\bibnamefont {Feng}}, \bibinfo
  {author} {\bibfnamefont {D.}~\bibnamefont {Wang}}, \bibinfo {author}
  {\bibfnamefont {S.}~\bibnamefont {Peng}}, \bibinfo {author} {\bibfnamefont
  {M.}~\bibnamefont {Li}}, \bibinfo {author} {\bibfnamefont {H.}~\bibnamefont
  {Wang}}, \bibinfo {author} {\bibfnamefont {Y.}~\bibnamefont {Xu}}, \bibinfo
  {author} {\bibfnamefont {T.}~\bibnamefont {Zhao}}, \bibinfo {author}
  {\bibfnamefont {B.}~\bibnamefont {Zhao}}, \bibinfo {author} {\bibfnamefont
  {S.}~\bibnamefont {Jiang}}, \emph {et~al.},\ }\bibfield  {title} {\enquote
  {\bibinfo {title} {{Pressure-induced structural evolution with suppression of
  the charge density wave state and dimensional crossover in CeTe 3}},}\
  }\href@noop {} {\bibfield  {journal} {\bibinfo  {journal} {Physical Review
  B}\ }\textbf {\bibinfo {volume} {109}},\ \bibinfo {pages} {094119} (\bibinfo
  {year} {2024})}\BibitemShut {NoStop}%
\bibitem [{\citenamefont {Zocco}\ \emph {et~al.}(2015)\citenamefont {Zocco},
  \citenamefont {Hamlin}, \citenamefont {Grube}, \citenamefont {Chu},
  \citenamefont {Kuo}, \citenamefont {Fisher},\ and\ \citenamefont
  {Maple}}]{Zocco2015}%
  \BibitemOpen
  \bibfield  {author} {\bibinfo {author} {\bibfnamefont {D.~A.}\ \bibnamefont
  {Zocco}}, \bibinfo {author} {\bibfnamefont {J.~J.}\ \bibnamefont {Hamlin}},
  \bibinfo {author} {\bibfnamefont {K.}~\bibnamefont {Grube}}, \bibinfo
  {author} {\bibfnamefont {J.-H.}\ \bibnamefont {Chu}}, \bibinfo {author}
  {\bibfnamefont {H.-H.}\ \bibnamefont {Kuo}}, \bibinfo {author} {\bibfnamefont
  {I.~R.}\ \bibnamefont {Fisher}},\ and\ \bibinfo {author} {\bibfnamefont
  {M.~B.}\ \bibnamefont {Maple}},\ }\bibfield  {title} {\enquote {\bibinfo
  {title} {{Pressure dependence of the charge-density-wave and superconducting
  states in ${\text{GdTe}}_{3}, {\text{TbTe}}_{3}$, and
  ${\text{DyTe}}_{3}$}},}\ }\href {https://doi.org/10.1103/PhysRevB.91.205114}
  {\bibfield  {journal} {\bibinfo  {journal} {Phys. Rev. B}\ }\textbf {\bibinfo
  {volume} {91}},\ \bibinfo {pages} {205114} (\bibinfo {year}
  {2015})}\BibitemShut {NoStop}%
\bibitem [{\citenamefont {Kopaczek}\ \emph {et~al.}(2022)\citenamefont
  {Kopaczek}, \citenamefont {Li}, \citenamefont {Yumigeta}, \citenamefont
  {Sailus}, \citenamefont {Sayyad}, \citenamefont {Moosavy}, \citenamefont
  {Kudrawiec},\ and\ \citenamefont {Tongay}}]{kopaczek2022}%
  \BibitemOpen
  \bibfield  {author} {\bibinfo {author} {\bibfnamefont {J.}~\bibnamefont
  {Kopaczek}}, \bibinfo {author} {\bibfnamefont {H.}~\bibnamefont {Li}},
  \bibinfo {author} {\bibfnamefont {K.}~\bibnamefont {Yumigeta}}, \bibinfo
  {author} {\bibfnamefont {R.}~\bibnamefont {Sailus}}, \bibinfo {author}
  {\bibfnamefont {M.~Y.}\ \bibnamefont {Sayyad}}, \bibinfo {author}
  {\bibfnamefont {S.~T.~R.}\ \bibnamefont {Moosavy}}, \bibinfo {author}
  {\bibfnamefont {R.}~\bibnamefont {Kudrawiec}},\ and\ \bibinfo {author}
  {\bibfnamefont {S.}~\bibnamefont {Tongay}},\ }\bibfield  {title} {\enquote
  {\bibinfo {title} {Pressure-induced suppression of charge density phases
  across the entire rare-earth tritellurides by optical spectroscopy},}\
  }\href@noop {} {\bibfield  {journal} {\bibinfo  {journal} {Journal of
  Materials Chemistry C}\ }\textbf {\bibinfo {volume} {10}},\ \bibinfo {pages}
  {11995--12000} (\bibinfo {year} {2022})}\BibitemShut {NoStop}%
\bibitem [{\citenamefont {Yumigeta}\ \emph {et~al.}(2022)\citenamefont
  {Yumigeta}, \citenamefont {Attarde}, \citenamefont {Kopaczek}, \citenamefont
  {Sayyad}, \citenamefont {Shen}, \citenamefont {Blei}, \citenamefont
  {Rajaei~Moosavy}, \citenamefont {Qin}, \citenamefont {Sailus},\ and\
  \citenamefont {Tongay}}]{yumigeta2022}%
  \BibitemOpen
  \bibfield  {author} {\bibinfo {author} {\bibfnamefont {K.}~\bibnamefont
  {Yumigeta}}, \bibinfo {author} {\bibfnamefont {Y.}~\bibnamefont {Attarde}},
  \bibinfo {author} {\bibfnamefont {J.}~\bibnamefont {Kopaczek}}, \bibinfo
  {author} {\bibfnamefont {M.~Y.}\ \bibnamefont {Sayyad}}, \bibinfo {author}
  {\bibfnamefont {Y.}~\bibnamefont {Shen}}, \bibinfo {author} {\bibfnamefont
  {M.}~\bibnamefont {Blei}}, \bibinfo {author} {\bibfnamefont {S.~T.}\
  \bibnamefont {Rajaei~Moosavy}}, \bibinfo {author} {\bibfnamefont
  {Y.}~\bibnamefont {Qin}}, \bibinfo {author} {\bibfnamefont {R.}~\bibnamefont
  {Sailus}},\ and\ \bibinfo {author} {\bibfnamefont {S.}~\bibnamefont
  {Tongay}},\ }\bibfield  {title} {\enquote {\bibinfo {title} {The phononic and
  charge density wave behavior of entire rare-earth tritelluride series with
  chemical pressure and temperature},}\ }\href@noop {} {\bibfield  {journal}
  {\bibinfo  {journal} {APL Materials}\ }\textbf {\bibinfo {volume} {10}}
  (\bibinfo {year} {2022})}\BibitemShut {NoStop}%
\bibitem [{\citenamefont {Saunot}\ \emph {et~al.}(2026)\citenamefont {Saunot},
  \citenamefont {Mikšić~Trontl}, \citenamefont {Klimovskikh}, \citenamefont
  {Vyalikh}, \citenamefont {Louat}, \citenamefont {Cacho}, \citenamefont
  {Kundu}, \citenamefont {Vescovo}, \citenamefont {Vobornik}, \citenamefont
  {Fedorov}, \citenamefont {Petrovic},\ and\ \citenamefont
  {Valla}}]{Saunot2026}%
  \BibitemOpen
  \bibfield  {author} {\bibinfo {author} {\bibfnamefont {A.}~\bibnamefont
  {Saunot}}, \bibinfo {author} {\bibfnamefont {V.}~\bibnamefont
  {Mikšić~Trontl}}, \bibinfo {author} {\bibfnamefont {I.~I.}\ \bibnamefont
  {Klimovskikh}}, \bibinfo {author} {\bibfnamefont {D.~V.}\ \bibnamefont
  {Vyalikh}}, \bibinfo {author} {\bibfnamefont {A.}~\bibnamefont {Louat}},
  \bibinfo {author} {\bibfnamefont {C.}~\bibnamefont {Cacho}}, \bibinfo
  {author} {\bibfnamefont {A.~K.}\ \bibnamefont {Kundu}}, \bibinfo {author}
  {\bibfnamefont {E.}~\bibnamefont {Vescovo}}, \bibinfo {author} {\bibfnamefont
  {I.}~\bibnamefont {Vobornik}}, \bibinfo {author} {\bibfnamefont
  {A.}~\bibnamefont {Fedorov}}, \bibinfo {author} {\bibfnamefont
  {C.}~\bibnamefont {Petrovic}},\ and\ \bibinfo {author} {\bibfnamefont
  {T.}~\bibnamefont {Valla}},\ }\bibfield  {title} {\enquote {\bibinfo {title}
  {{Interplay of Kondo physics with incommensurate charge density waves in
  CeTe3}},}\ }\href {https://doi.org/10.1063/5.0313165} {\bibfield  {journal}
  {\bibinfo  {journal} {Applied Physics Letters}\ }\textbf {\bibinfo {volume}
  {128}},\ \bibinfo {pages} {121909} (\bibinfo {year} {2026})}\BibitemShut
  {NoStop}%
\bibitem [{\citenamefont {Ru}\ and\ \citenamefont {Fisher}(2006)}]{ru2006}%
  \BibitemOpen
  \bibfield  {author} {\bibinfo {author} {\bibfnamefont {N.}~\bibnamefont
  {Ru}}\ and\ \bibinfo {author} {\bibfnamefont {I.~R.}\ \bibnamefont
  {Fisher}},\ }\bibfield  {title} {\enquote {\bibinfo {title} {Thermodynamic
  and transport properties of $\mathrm{Y}{\mathrm{te}}_{3}$,
  $\mathrm{La}{\mathrm{te}}_{3}$, and $\mathrm{Ce}{\mathrm{te}}_{3}$},}\ }\href
  {https://doi.org/10.1103/PhysRevB.73.033101} {\bibfield  {journal} {\bibinfo
  {journal} {Phys. Rev. B}\ }\textbf {\bibinfo {volume} {73}},\ \bibinfo
  {pages} {033101} (\bibinfo {year} {2006})}\BibitemShut {NoStop}%
\bibitem [{\citenamefont {Schmitt}\ \emph
  {et~al.}(2011{\natexlab{b}})\citenamefont {Schmitt}, \citenamefont
  {Kirchmann}, \citenamefont {Bovensiepen}, \citenamefont {Moore},
  \citenamefont {Chu}, \citenamefont {Lu}, \citenamefont {Rettig},
  \citenamefont {Wolf}, \citenamefont {Fisher},\ and\ \citenamefont
  {Shen}}]{Schmitt2011a}%
  \BibitemOpen
  \bibfield  {author} {\bibinfo {author} {\bibfnamefont {F.}~\bibnamefont
  {Schmitt}}, \bibinfo {author} {\bibfnamefont {P.~S.}\ \bibnamefont
  {Kirchmann}}, \bibinfo {author} {\bibfnamefont {U.}~\bibnamefont
  {Bovensiepen}}, \bibinfo {author} {\bibfnamefont {R.~G.}\ \bibnamefont
  {Moore}}, \bibinfo {author} {\bibfnamefont {J.-H.}\ \bibnamefont {Chu}},
  \bibinfo {author} {\bibfnamefont {D.~H.}\ \bibnamefont {Lu}}, \bibinfo
  {author} {\bibfnamefont {L.}~\bibnamefont {Rettig}}, \bibinfo {author}
  {\bibfnamefont {M.}~\bibnamefont {Wolf}}, \bibinfo {author} {\bibfnamefont
  {I.~R.}\ \bibnamefont {Fisher}},\ and\ \bibinfo {author} {\bibfnamefont
  {Z.-X.}\ \bibnamefont {Shen}},\ }\bibfield  {title} {\enquote {\bibinfo
  {title} {{Ultrafast electron dynamics in the charge density wave material
  TbTe3}},}\ }\href {https://doi.org/10.1088/1367-2630/13/6/063022} {\bibfield
  {journal} {\bibinfo  {journal} {New Journal of Physics}\ }\textbf {\bibinfo
  {volume} {13}},\ \bibinfo {pages} {063022} (\bibinfo {year}
  {2011}{\natexlab{b}})}\BibitemShut {NoStop}%
\bibitem [{\citenamefont {Rettig}\ \emph {et~al.}(2014)\citenamefont {Rettig},
  \citenamefont {Chu}, \citenamefont {Fisher}, \citenamefont {Bovensiepen},\
  and\ \citenamefont {Wolf}}]{rettig2014}%
  \BibitemOpen
  \bibfield  {author} {\bibinfo {author} {\bibfnamefont {L.}~\bibnamefont
  {Rettig}}, \bibinfo {author} {\bibfnamefont {J.-H.}\ \bibnamefont {Chu}},
  \bibinfo {author} {\bibfnamefont {I.}~\bibnamefont {Fisher}}, \bibinfo
  {author} {\bibfnamefont {U.}~\bibnamefont {Bovensiepen}},\ and\ \bibinfo
  {author} {\bibfnamefont {M.}~\bibnamefont {Wolf}},\ }\bibfield  {title}
  {\enquote {\bibinfo {title} {Coherent dynamics of the charge density wave gap
  in tritellurides},}\ }\href@noop {} {\bibfield  {journal} {\bibinfo
  {journal} {Faraday discussions}\ }\textbf {\bibinfo {volume} {171}},\
  \bibinfo {pages} {299--310} (\bibinfo {year} {2014})}\BibitemShut {NoStop}%
\bibitem [{\citenamefont {Yusupov}\ \emph {et~al.}(2010)\citenamefont
  {Yusupov}, \citenamefont {Mertelj}, \citenamefont {Kabanov}, \citenamefont
  {Brazovskii}, \citenamefont {Kusar}, \citenamefont {Chu}, \citenamefont
  {Fisher},\ and\ \citenamefont {Mihailovic}}]{yusupov2010}%
  \BibitemOpen
  \bibfield  {author} {\bibinfo {author} {\bibfnamefont {R.}~\bibnamefont
  {Yusupov}}, \bibinfo {author} {\bibfnamefont {T.}~\bibnamefont {Mertelj}},
  \bibinfo {author} {\bibfnamefont {V.~V.}\ \bibnamefont {Kabanov}}, \bibinfo
  {author} {\bibfnamefont {S.}~\bibnamefont {Brazovskii}}, \bibinfo {author}
  {\bibfnamefont {P.}~\bibnamefont {Kusar}}, \bibinfo {author} {\bibfnamefont
  {J.-H.}\ \bibnamefont {Chu}}, \bibinfo {author} {\bibfnamefont {I.~R.}\
  \bibnamefont {Fisher}},\ and\ \bibinfo {author} {\bibfnamefont
  {D.}~\bibnamefont {Mihailovic}},\ }\bibfield  {title} {\enquote {\bibinfo
  {title} {Coherent dynamics of macroscopic electronic order through a symmetry
  breaking transition},}\ }\href@noop {} {\bibfield  {journal} {\bibinfo
  {journal} {Nature Physics}\ }\textbf {\bibinfo {volume} {6}},\ \bibinfo
  {pages} {681--684} (\bibinfo {year} {2010})}\BibitemShut {NoStop}%
\bibitem [{\citenamefont {Zong}\ \emph {et~al.}(2019)\citenamefont {Zong},
  \citenamefont {Kogar}, \citenamefont {Bie}, \citenamefont {Rohwer},
  \citenamefont {Lee}, \citenamefont {Baldini}, \citenamefont {Erge{\c{c}}en},
  \citenamefont {Yilmaz}, \citenamefont {Freelon}, \citenamefont {Sie} \emph
  {et~al.}}]{zong2019}%
  \BibitemOpen
  \bibfield  {author} {\bibinfo {author} {\bibfnamefont {A.}~\bibnamefont
  {Zong}}, \bibinfo {author} {\bibfnamefont {A.}~\bibnamefont {Kogar}},
  \bibinfo {author} {\bibfnamefont {Y.-Q.}\ \bibnamefont {Bie}}, \bibinfo
  {author} {\bibfnamefont {T.}~\bibnamefont {Rohwer}}, \bibinfo {author}
  {\bibfnamefont {C.}~\bibnamefont {Lee}}, \bibinfo {author} {\bibfnamefont
  {E.}~\bibnamefont {Baldini}}, \bibinfo {author} {\bibfnamefont
  {E.}~\bibnamefont {Erge{\c{c}}en}}, \bibinfo {author} {\bibfnamefont {M.~B.}\
  \bibnamefont {Yilmaz}}, \bibinfo {author} {\bibfnamefont {B.}~\bibnamefont
  {Freelon}}, \bibinfo {author} {\bibfnamefont {E.~J.}\ \bibnamefont {Sie}},
  \emph {et~al.},\ }\bibfield  {title} {\enquote {\bibinfo {title} {Evidence
  for topological defects in a photoinduced phase transition},}\ }\href@noop {}
  {\bibfield  {journal} {\bibinfo  {journal} {Nature Physics}\ }\textbf
  {\bibinfo {volume} {15}},\ \bibinfo {pages} {27--31} (\bibinfo {year}
  {2019})}\BibitemShut {NoStop}%
\bibitem [{\citenamefont {Kogar}\ \emph {et~al.}(2020)\citenamefont {Kogar},
  \citenamefont {Zong}, \citenamefont {Dolgirev}, \citenamefont {Shen},
  \citenamefont {Straquadine}, \citenamefont {Bie}, \citenamefont {Wang},
  \citenamefont {Rohwer}, \citenamefont {Tung}, \citenamefont {Yang} \emph
  {et~al.}}]{kogar2020}%
  \BibitemOpen
  \bibfield  {author} {\bibinfo {author} {\bibfnamefont {A.}~\bibnamefont
  {Kogar}}, \bibinfo {author} {\bibfnamefont {A.}~\bibnamefont {Zong}},
  \bibinfo {author} {\bibfnamefont {P.~E.}\ \bibnamefont {Dolgirev}}, \bibinfo
  {author} {\bibfnamefont {X.}~\bibnamefont {Shen}}, \bibinfo {author}
  {\bibfnamefont {J.}~\bibnamefont {Straquadine}}, \bibinfo {author}
  {\bibfnamefont {Y.-Q.}\ \bibnamefont {Bie}}, \bibinfo {author} {\bibfnamefont
  {X.}~\bibnamefont {Wang}}, \bibinfo {author} {\bibfnamefont {T.}~\bibnamefont
  {Rohwer}}, \bibinfo {author} {\bibfnamefont {I.-C.}\ \bibnamefont {Tung}},
  \bibinfo {author} {\bibfnamefont {Y.}~\bibnamefont {Yang}}, \emph {et~al.},\
  }\bibfield  {title} {\enquote {\bibinfo {title} {{Light-induced charge
  density wave in LaTe3}},}\ }\href@noop {} {\bibfield  {journal} {\bibinfo
  {journal} {Nature Physics}\ }\textbf {\bibinfo {volume} {16}},\ \bibinfo
  {pages} {159--163} (\bibinfo {year} {2020})}\BibitemShut {NoStop}%
\bibitem [{\citenamefont {Trigo}\ \emph {et~al.}(2021)\citenamefont {Trigo},
  \citenamefont {Giraldo-Gallo}, \citenamefont {Clark}, \citenamefont {Kozina},
  \citenamefont {Henighan}, \citenamefont {Jiang}, \citenamefont {Chollet},
  \citenamefont {Fisher}, \citenamefont {Glownia}, \citenamefont {Katayama},
  \citenamefont {Kirchmann}, \citenamefont {Leuenberger}, \citenamefont {Liu},
  \citenamefont {Reis}, \citenamefont {Shen},\ and\ \citenamefont
  {Zhu}}]{trigo2021}%
  \BibitemOpen
  \bibfield  {author} {\bibinfo {author} {\bibfnamefont {M.}~\bibnamefont
  {Trigo}}, \bibinfo {author} {\bibfnamefont {P.}~\bibnamefont
  {Giraldo-Gallo}}, \bibinfo {author} {\bibfnamefont {J.~N.}\ \bibnamefont
  {Clark}}, \bibinfo {author} {\bibfnamefont {M.~E.}\ \bibnamefont {Kozina}},
  \bibinfo {author} {\bibfnamefont {T.}~\bibnamefont {Henighan}}, \bibinfo
  {author} {\bibfnamefont {M.~P.}\ \bibnamefont {Jiang}}, \bibinfo {author}
  {\bibfnamefont {M.}~\bibnamefont {Chollet}}, \bibinfo {author} {\bibfnamefont
  {I.~R.}\ \bibnamefont {Fisher}}, \bibinfo {author} {\bibfnamefont {J.~M.}\
  \bibnamefont {Glownia}}, \bibinfo {author} {\bibfnamefont {T.}~\bibnamefont
  {Katayama}}, \bibinfo {author} {\bibfnamefont {P.~S.}\ \bibnamefont
  {Kirchmann}}, \bibinfo {author} {\bibfnamefont {D.}~\bibnamefont
  {Leuenberger}}, \bibinfo {author} {\bibfnamefont {H.}~\bibnamefont {Liu}},
  \bibinfo {author} {\bibfnamefont {D.~A.}\ \bibnamefont {Reis}}, \bibinfo
  {author} {\bibfnamefont {Z.~X.}\ \bibnamefont {Shen}},\ and\ \bibinfo
  {author} {\bibfnamefont {D.}~\bibnamefont {Zhu}},\ }\bibfield  {title}
  {\enquote {\bibinfo {title} {Ultrafast formation of domain walls of a charge
  density wave in ${\mathrm{smte}}_{3}$},}\ }\href
  {https://doi.org/10.1103/PhysRevB.103.054109} {\bibfield  {journal} {\bibinfo
   {journal} {Phys. Rev. B}\ }\textbf {\bibinfo {volume} {103}},\ \bibinfo
  {pages} {054109} (\bibinfo {year} {2021})}\BibitemShut {NoStop}%
\bibitem [{\citenamefont {Zhou}\ \emph {et~al.}(2021)\citenamefont {Zhou},
  \citenamefont {Williams}, \citenamefont {Sun}, \citenamefont {Malliakas},
  \citenamefont {Kanatzidis}, \citenamefont {Kemper},\ and\ \citenamefont
  {Ruan}}]{zhou2021}%
  \BibitemOpen
  \bibfield  {author} {\bibinfo {author} {\bibfnamefont {F.}~\bibnamefont
  {Zhou}}, \bibinfo {author} {\bibfnamefont {J.}~\bibnamefont {Williams}},
  \bibinfo {author} {\bibfnamefont {S.}~\bibnamefont {Sun}}, \bibinfo {author}
  {\bibfnamefont {C.~D.}\ \bibnamefont {Malliakas}}, \bibinfo {author}
  {\bibfnamefont {M.~G.}\ \bibnamefont {Kanatzidis}}, \bibinfo {author}
  {\bibfnamefont {A.~F.}\ \bibnamefont {Kemper}},\ and\ \bibinfo {author}
  {\bibfnamefont {C.-Y.}\ \bibnamefont {Ruan}},\ }\bibfield  {title} {\enquote
  {\bibinfo {title} {Nonequilibrium dynamics of spontaneous symmetry breaking
  into a hidden state of charge-density wave},}\ }\href@noop {} {\bibfield
  {journal} {\bibinfo  {journal} {Nature communications}\ }\textbf {\bibinfo
  {volume} {12}},\ \bibinfo {pages} {566} (\bibinfo {year} {2021})}\BibitemShut
  {NoStop}%
\bibitem [{\citenamefont {Sch{\"a}fer}, \citenamefont {Kabanov},\ and\
  \citenamefont {Demsar}(2014)}]{schaefer2014}%
  \BibitemOpen
  \bibfield  {author} {\bibinfo {author} {\bibfnamefont {H.}~\bibnamefont
  {Sch{\"a}fer}}, \bibinfo {author} {\bibfnamefont {V.~V.}\ \bibnamefont
  {Kabanov}},\ and\ \bibinfo {author} {\bibfnamefont {J.}~\bibnamefont
  {Demsar}},\ }\bibfield  {title} {\enquote {\bibinfo {title} {Collective modes
  in quasi-one-dimensional charge-density wave systems probed by femtosecond
  time-resolved optical studies},}\ }\href@noop {} {\bibfield  {journal}
  {\bibinfo  {journal} {Physical Review B}\ }\textbf {\bibinfo {volume} {89}},\
  \bibinfo {pages} {045106} (\bibinfo {year} {2014})}\BibitemShut {NoStop}%
\bibitem [{\citenamefont {Sagar}\ \emph {et~al.}(2008)\citenamefont {Sagar},
  \citenamefont {Fausti}, \citenamefont {Yue}, \citenamefont {Kuntscher},
  \citenamefont {van Smaalen},\ and\ \citenamefont
  {Van~Loosdrecht}}]{sagar2008}%
  \BibitemOpen
  \bibfield  {author} {\bibinfo {author} {\bibfnamefont {D.}~\bibnamefont
  {Sagar}}, \bibinfo {author} {\bibfnamefont {D.}~\bibnamefont {Fausti}},
  \bibinfo {author} {\bibfnamefont {S.}~\bibnamefont {Yue}}, \bibinfo {author}
  {\bibfnamefont {C.~A.}\ \bibnamefont {Kuntscher}}, \bibinfo {author}
  {\bibfnamefont {S.}~\bibnamefont {van Smaalen}},\ and\ \bibinfo {author}
  {\bibfnamefont {P.}~\bibnamefont {Van~Loosdrecht}},\ }\bibfield  {title}
  {\enquote {\bibinfo {title} {{A Raman study of the charge-density-wave state
  in A0. 3MoO3 (A= K, Rb)}},}\ }\href@noop {} {\bibfield  {journal} {\bibinfo
  {journal} {New Journal of Physics}\ }\textbf {\bibinfo {volume} {10}},\
  \bibinfo {pages} {023043} (\bibinfo {year} {2008})}\BibitemShut {NoStop}%
\bibitem [{\citenamefont {Hansen}\ \emph {et~al.}(2023)\citenamefont {Hansen},
  \citenamefont {Palan}, \citenamefont {Hahn}, \citenamefont {Thomson},
  \citenamefont {Warawa}, \citenamefont {Roskos}, \citenamefont {Demsar},
  \citenamefont {Pientka}, \citenamefont {Tsyplyatyev},\ and\ \citenamefont
  {Kopietz}}]{hansen2023}%
  \BibitemOpen
  \bibfield  {author} {\bibinfo {author} {\bibfnamefont {M.~O.}\ \bibnamefont
  {Hansen}}, \bibinfo {author} {\bibfnamefont {Y.}~\bibnamefont {Palan}},
  \bibinfo {author} {\bibfnamefont {V.}~\bibnamefont {Hahn}}, \bibinfo {author}
  {\bibfnamefont {M.~D.}\ \bibnamefont {Thomson}}, \bibinfo {author}
  {\bibfnamefont {K.}~\bibnamefont {Warawa}}, \bibinfo {author} {\bibfnamefont
  {H.~G.}\ \bibnamefont {Roskos}}, \bibinfo {author} {\bibfnamefont
  {J.}~\bibnamefont {Demsar}}, \bibinfo {author} {\bibfnamefont
  {F.}~\bibnamefont {Pientka}}, \bibinfo {author} {\bibfnamefont
  {O.}~\bibnamefont {Tsyplyatyev}},\ and\ \bibinfo {author} {\bibfnamefont
  {P.}~\bibnamefont {Kopietz}},\ }\bibfield  {title} {\enquote {\bibinfo
  {title} {{Collective modes in the charge density wave state of K0.3MoO3: Role
  of long-range Coulomb interactions revisited}},}\ }\href@noop {} {\bibfield
  {journal} {\bibinfo  {journal} {Physical Review B}\ }\textbf {\bibinfo
  {volume} {108}},\ \bibinfo {pages} {045148} (\bibinfo {year}
  {2023})}\BibitemShut {NoStop}%
\bibitem [{\citenamefont {Rice}(1978)}]{Rice1978}%
  \BibitemOpen
  \bibfield  {author} {\bibinfo {author} {\bibfnamefont {M.}~\bibnamefont
  {Rice}},\ }\bibfield  {title} {\enquote {\bibinfo {title} {{Dynamical
  properties of the Peierls-Fröhlich state on the many-phonon-coupling
  model}},}\ }\href
  {https://doi.org/https://doi.org/10.1016/0038-1098(78)90912-2} {\bibfield
  {journal} {\bibinfo  {journal} {Solid State Communications}\ }\textbf
  {\bibinfo {volume} {25}},\ \bibinfo {pages} {1083--1086} (\bibinfo {year}
  {1978})}\BibitemShut {NoStop}%
\bibitem [{\citenamefont {Thomson}\ \emph {et~al.}(2017)\citenamefont
  {Thomson}, \citenamefont {Rabia}, \citenamefont {Meng},\ and\ \citenamefont
  {Roskos}}]{thomson2017}%
  \BibitemOpen
  \bibfield  {author} {\bibinfo {author} {\bibfnamefont {M.}~\bibnamefont
  {Thomson}}, \bibinfo {author} {\bibfnamefont {K.}~\bibnamefont {Rabia}},
  \bibinfo {author} {\bibfnamefont {F.}~\bibnamefont {Meng}},\ and\ \bibinfo
  {author} {\bibfnamefont {H.}~\bibnamefont {Roskos}},\ }\bibfield  {title}
  {\enquote {\bibinfo {title} {Phase-channel dynamics reveal the role of
  impurities and screening in a quasi-one-dimensional charge-density wave
  system},}\ }\href@noop {} {\bibfield  {journal} {\bibinfo  {journal}
  {Scientific Reports}\ }\textbf {\bibinfo {volume} {7}},\ \bibinfo {pages}
  {2039} (\bibinfo {year} {2017})}\BibitemShut {NoStop}%
\bibitem [{\citenamefont {Demsar}\ \emph {et~al.}(2006)\citenamefont {Demsar},
  \citenamefont {Thorsm{\o}lle}, \citenamefont {Sarrao},\ and\ \citenamefont
  {Taylor}}]{demsar2006a}%
  \BibitemOpen
  \bibfield  {author} {\bibinfo {author} {\bibfnamefont {J.}~\bibnamefont
  {Demsar}}, \bibinfo {author} {\bibfnamefont {V.~K.}\ \bibnamefont
  {Thorsm{\o}lle}}, \bibinfo {author} {\bibfnamefont {J.~L.}\ \bibnamefont
  {Sarrao}},\ and\ \bibinfo {author} {\bibfnamefont {A.~J.}\ \bibnamefont
  {Taylor}},\ }\bibfield  {title} {\enquote {\bibinfo {title} {{Photoexcited
  electron dynamics in Kondo insulators and heavy fermions}},}\ }\href@noop {}
  {\bibfield  {journal} {\bibinfo  {journal} {Physical review letters}\
  }\textbf {\bibinfo {volume} {96}},\ \bibinfo {pages} {037401} (\bibinfo
  {year} {2006})}\BibitemShut {NoStop}%
\bibitem [{\citenamefont {Demsar}, \citenamefont {Sarrao},\ and\ \citenamefont
  {Taylor}(2006)}]{demsar2006b}%
  \BibitemOpen
  \bibfield  {author} {\bibinfo {author} {\bibfnamefont {J.}~\bibnamefont
  {Demsar}}, \bibinfo {author} {\bibfnamefont {J.~L.}\ \bibnamefont {Sarrao}},\
  and\ \bibinfo {author} {\bibfnamefont {A.~J.}\ \bibnamefont {Taylor}},\
  }\bibfield  {title} {\enquote {\bibinfo {title} {Dynamics of photoexcited
  quasiparticles in heavy electron compounds},}\ }\href@noop {} {\bibfield
  {journal} {\bibinfo  {journal} {Journal of Physics: Condensed Matter}\
  }\textbf {\bibinfo {volume} {18}},\ \bibinfo {pages} {R281--R314} (\bibinfo
  {year} {2006})}\BibitemShut {NoStop}%
\bibitem [{\citenamefont {Chen}\ \emph {et~al.}(2014)\citenamefont {Chen},
  \citenamefont {Hu}, \citenamefont {Dong},\ and\ \citenamefont
  {Wang}}]{chen2014}%
  \BibitemOpen
  \bibfield  {author} {\bibinfo {author} {\bibfnamefont {R.~Y.}\ \bibnamefont
  {Chen}}, \bibinfo {author} {\bibfnamefont {B.~F.}\ \bibnamefont {Hu}},
  \bibinfo {author} {\bibfnamefont {T.}~\bibnamefont {Dong}},\ and\ \bibinfo
  {author} {\bibfnamefont {N.~L.}\ \bibnamefont {Wang}},\ }\bibfield  {title}
  {\enquote {\bibinfo {title} {Revealing multiple charge-density-wave orders in
  ${\text{tbte}}_{3}$ by optical conductivity and ultrafast pump-probe
  experiments},}\ }\href {https://doi.org/10.1103/PhysRevB.89.075114}
  {\bibfield  {journal} {\bibinfo  {journal} {Phys. Rev. B}\ }\textbf {\bibinfo
  {volume} {89}},\ \bibinfo {pages} {075114} (\bibinfo {year}
  {2014})}\BibitemShut {NoStop}%
\bibitem [{\citenamefont {Tsuchiya}\ \emph {et~al.}(2015)\citenamefont
  {Tsuchiya}, \citenamefont {Sugawara}, \citenamefont {Tanda},\ and\
  \citenamefont {Toda}}]{tsuchiya2015}%
  \BibitemOpen
  \bibfield  {author} {\bibinfo {author} {\bibfnamefont {S.}~\bibnamefont
  {Tsuchiya}}, \bibinfo {author} {\bibfnamefont {Y.}~\bibnamefont {Sugawara}},
  \bibinfo {author} {\bibfnamefont {S.}~\bibnamefont {Tanda}},\ and\ \bibinfo
  {author} {\bibfnamefont {Y.}~\bibnamefont {Toda}},\ }\bibfield  {title}
  {\enquote {\bibinfo {title} {Symmetry-dependent carrier relaxation dynamics
  and charge--density--wave transition in dyte3 probed by polarized femtosecond
  spectroscopy},}\ }\href@noop {} {\bibfield  {journal} {\bibinfo  {journal}
  {Journal of optics}\ }\textbf {\bibinfo {volume} {17}},\ \bibinfo {pages}
  {085501} (\bibinfo {year} {2015})}\BibitemShut {NoStop}%
\bibitem [{\citenamefont {Demsar}\ \emph {et~al.}(2002)\citenamefont {Demsar},
  \citenamefont {Forr{\'o}}, \citenamefont {Berger},\ and\ \citenamefont
  {Mihailovic}}]{demsar2002}%
  \BibitemOpen
  \bibfield  {author} {\bibinfo {author} {\bibfnamefont {J.}~\bibnamefont
  {Demsar}}, \bibinfo {author} {\bibfnamefont {L.}~\bibnamefont {Forr{\'o}}},
  \bibinfo {author} {\bibfnamefont {H.}~\bibnamefont {Berger}},\ and\ \bibinfo
  {author} {\bibfnamefont {D.}~\bibnamefont {Mihailovic}},\ }\bibfield  {title}
  {\enquote {\bibinfo {title} {{Femtosecond snapshots of gap-forming
  charge-density-wave correlations in quasi-two-dimensional dichalcogenides 1
  T-TaS2 and 2H-TaSe2}},}\ }\href@noop {} {\bibfield  {journal} {\bibinfo
  {journal} {Physical review B}\ }\textbf {\bibinfo {volume} {66}},\ \bibinfo
  {pages} {041101} (\bibinfo {year} {2002})}\BibitemShut {NoStop}%
\bibitem [{\citenamefont {Yue}\ \emph {et~al.}(2023)\citenamefont {Yue},
  \citenamefont {Pokharel}, \citenamefont {Demsar}, \citenamefont {Zhang},
  \citenamefont {Li}, \citenamefont {Dong},\ and\ \citenamefont
  {Wang}}]{Li2023}%
  \BibitemOpen
  \bibfield  {author} {\bibinfo {author} {\bibfnamefont {L.}~\bibnamefont
  {Yue}}, \bibinfo {author} {\bibfnamefont {A.~R.}\ \bibnamefont {Pokharel}},
  \bibinfo {author} {\bibfnamefont {J.}~\bibnamefont {Demsar}}, \bibinfo
  {author} {\bibfnamefont {S.}~\bibnamefont {Zhang}}, \bibinfo {author}
  {\bibfnamefont {Y.}~\bibnamefont {Li}}, \bibinfo {author} {\bibfnamefont
  {T.}~\bibnamefont {Dong}},\ and\ \bibinfo {author} {\bibfnamefont
  {N.}~\bibnamefont {Wang}},\ }\bibfield  {title} {\enquote {\bibinfo {title}
  {{Highly anisotropic transient optical response of charge density wave order
  in ${\mathrm{ZrTe}}_{3}$}},}\ }\href
  {https://doi.org/10.1103/PhysRevB.107.165115} {\bibfield  {journal} {\bibinfo
   {journal} {Phys. Rev. B}\ }\textbf {\bibinfo {volume} {107}},\ \bibinfo
  {pages} {165115} (\bibinfo {year} {2023})}\BibitemShut {NoStop}%
\bibitem [{\citenamefont {Groeneveld}, \citenamefont {Sprik},\ and\
  \citenamefont {Lagendijk}(1995)}]{groeneveld1995}%
  \BibitemOpen
  \bibfield  {author} {\bibinfo {author} {\bibfnamefont {R.~H.}\ \bibnamefont
  {Groeneveld}}, \bibinfo {author} {\bibfnamefont {R.}~\bibnamefont {Sprik}},\
  and\ \bibinfo {author} {\bibfnamefont {A.}~\bibnamefont {Lagendijk}},\
  }\bibfield  {title} {\enquote {\bibinfo {title} {{Femtosecond spectroscopy of
  electron-electron and electron-phonon energy relaxation in Ag and Au}},}\
  }\href@noop {} {\bibfield  {journal} {\bibinfo  {journal} {Physical Review
  B}\ }\textbf {\bibinfo {volume} {51}},\ \bibinfo {pages} {11433} (\bibinfo
  {year} {1995})}\BibitemShut {NoStop}%
\bibitem [{\citenamefont {Ahn}\ \emph {et~al.}(2004)\citenamefont {Ahn},
  \citenamefont {Graf}, \citenamefont {Trugman}, \citenamefont {Demsar},
  \citenamefont {Averitt}, \citenamefont {Sarrao},\ and\ \citenamefont
  {Taylor}}]{Ahn2004}%
  \BibitemOpen
  \bibfield  {author} {\bibinfo {author} {\bibfnamefont {K.~H.}\ \bibnamefont
  {Ahn}}, \bibinfo {author} {\bibfnamefont {M.~J.}\ \bibnamefont {Graf}},
  \bibinfo {author} {\bibfnamefont {S.~A.}\ \bibnamefont {Trugman}}, \bibinfo
  {author} {\bibfnamefont {J.}~\bibnamefont {Demsar}}, \bibinfo {author}
  {\bibfnamefont {R.~D.}\ \bibnamefont {Averitt}}, \bibinfo {author}
  {\bibfnamefont {J.~L.}\ \bibnamefont {Sarrao}},\ and\ \bibinfo {author}
  {\bibfnamefont {A.~J.}\ \bibnamefont {Taylor}},\ }\bibfield  {title}
  {\enquote {\bibinfo {title} {Ultrafast quasiparticle relaxation dynamics in
  normal metals and heavy-fermion materials},}\ }\href
  {https://doi.org/10.1103/PhysRevB.69.045114} {\bibfield  {journal} {\bibinfo
  {journal} {Phys. Rev. B}\ }\textbf {\bibinfo {volume} {69}},\ \bibinfo
  {pages} {045114} (\bibinfo {year} {2004})}\BibitemShut {NoStop}%
\bibitem [{\citenamefont {Pokharel}\ \emph {et~al.}(2021)\citenamefont
  {Pokharel}, \citenamefont {Agustsson}, \citenamefont {Kabanov}, \citenamefont
  {Iga}, \citenamefont {Takabatake}, \citenamefont {Okamura},\ and\
  \citenamefont {Demsar}}]{pokharel2021}%
  \BibitemOpen
  \bibfield  {author} {\bibinfo {author} {\bibfnamefont {A.}~\bibnamefont
  {Pokharel}}, \bibinfo {author} {\bibfnamefont {S.}~\bibnamefont {Agustsson}},
  \bibinfo {author} {\bibfnamefont {V.}~\bibnamefont {Kabanov}}, \bibinfo
  {author} {\bibfnamefont {F.}~\bibnamefont {Iga}}, \bibinfo {author}
  {\bibfnamefont {T.}~\bibnamefont {Takabatake}}, \bibinfo {author}
  {\bibfnamefont {H.}~\bibnamefont {Okamura}},\ and\ \bibinfo {author}
  {\bibfnamefont {J.}~\bibnamefont {Demsar}},\ }\bibfield  {title} {\enquote
  {\bibinfo {title} {{Robust hybridization gap in the Kondo insulator YbB12
  probed by femtosecond optical spectroscopy}},}\ }\href@noop {} {\bibfield
  {journal} {\bibinfo  {journal} {Physical Review B}\ }\textbf {\bibinfo
  {volume} {103}},\ \bibinfo {pages} {115134} (\bibinfo {year}
  {2021})}\BibitemShut {NoStop}%
\bibitem [{\citenamefont {Tuvia}\ \emph {et~al.}(2026)\citenamefont {Tuvia},
  \citenamefont {Kang}, \citenamefont {Golovanova}, \citenamefont {Chen},
  \citenamefont {Wang}, \citenamefont {Ma}, \citenamefont {Liu}, \citenamefont
  {Grossman}, \citenamefont {Sung}, \citenamefont {Shotton} \emph
  {et~al.}}]{tuvia2026}%
  \BibitemOpen
  \bibfield  {author} {\bibinfo {author} {\bibfnamefont {G.}~\bibnamefont
  {Tuvia}}, \bibinfo {author} {\bibfnamefont {R.}~\bibnamefont {Kang}},
  \bibinfo {author} {\bibfnamefont {D.}~\bibnamefont {Golovanova}}, \bibinfo
  {author} {\bibfnamefont {Y.}~\bibnamefont {Chen}}, \bibinfo {author}
  {\bibfnamefont {Y.}~\bibnamefont {Wang}}, \bibinfo {author} {\bibfnamefont
  {Z.}~\bibnamefont {Ma}}, \bibinfo {author} {\bibfnamefont {M.}~\bibnamefont
  {Liu}}, \bibinfo {author} {\bibfnamefont {C.}~\bibnamefont {Grossman}},
  \bibinfo {author} {\bibfnamefont {S.~H.}\ \bibnamefont {Sung}}, \bibinfo
  {author} {\bibfnamefont {J.}~\bibnamefont {Shotton}}, \emph {et~al.},\
  }\bibfield  {title} {\enquote {\bibinfo {title} {{Kondo Reshapes Multiple
  Orders in a $5f$ van der Waals Material}},}\ }\href@noop {} {\bibfield
  {journal} {\bibinfo  {journal} {arXiv preprint arXiv:2602.22451}\ } (\bibinfo
  {year} {2026})}\BibitemShut {NoStop}%
\bibitem [{\citenamefont {Zeiger}\ \emph {et~al.}(1992)\citenamefont {Zeiger},
  \citenamefont {Vidal}, \citenamefont {Cheng}, \citenamefont {Ippen},
  \citenamefont {Dresselhaus},\ and\ \citenamefont {Dresselhaus}}]{zeiger1992}%
  \BibitemOpen
  \bibfield  {author} {\bibinfo {author} {\bibfnamefont {H.}~\bibnamefont
  {Zeiger}}, \bibinfo {author} {\bibfnamefont {J.}~\bibnamefont {Vidal}},
  \bibinfo {author} {\bibfnamefont {T.}~\bibnamefont {Cheng}}, \bibinfo
  {author} {\bibfnamefont {E.}~\bibnamefont {Ippen}}, \bibinfo {author}
  {\bibfnamefont {G.}~\bibnamefont {Dresselhaus}},\ and\ \bibinfo {author}
  {\bibfnamefont {M.}~\bibnamefont {Dresselhaus}},\ }\bibfield  {title}
  {\enquote {\bibinfo {title} {Theory for displacive excitation of coherent
  phonons},}\ }\href@noop {} {\bibfield  {journal} {\bibinfo  {journal}
  {Physical Review B}\ }\textbf {\bibinfo {volume} {45}},\ \bibinfo {pages}
  {768} (\bibinfo {year} {1992})}\BibitemShut {NoStop}%
\bibitem [{\citenamefont {Johannes}\ and\ \citenamefont
  {Mazin}(2008)}]{mazin2008}%
  \BibitemOpen
  \bibfield  {author} {\bibinfo {author} {\bibfnamefont {M.}~\bibnamefont
  {Johannes}}\ and\ \bibinfo {author} {\bibfnamefont {I.}~\bibnamefont
  {Mazin}},\ }\bibfield  {title} {\enquote {\bibinfo {title} {Fermi surface
  nesting and the origin of charge density waves in metals},}\ }\href@noop {}
  {\bibfield  {journal} {\bibinfo  {journal} {Physical Review B—Condensed
  Matter and Materials Physics}\ }\textbf {\bibinfo {volume} {77}},\ \bibinfo
  {pages} {165135} (\bibinfo {year} {2008})}\BibitemShut {NoStop}%
\bibitem [{\citenamefont {Ru}\ \emph {et~al.}(2008)\citenamefont {Ru},
  \citenamefont {Condron}, \citenamefont {Margulis}, \citenamefont {Shin},
  \citenamefont {Laverock}, \citenamefont {Dugdale}, \citenamefont {Toney},\
  and\ \citenamefont {Fisher}}]{ru2008}%
  \BibitemOpen
  \bibfield  {author} {\bibinfo {author} {\bibfnamefont {N.}~\bibnamefont
  {Ru}}, \bibinfo {author} {\bibfnamefont {C.}~\bibnamefont {Condron}},
  \bibinfo {author} {\bibfnamefont {G.}~\bibnamefont {Margulis}}, \bibinfo
  {author} {\bibfnamefont {K.}~\bibnamefont {Shin}}, \bibinfo {author}
  {\bibfnamefont {J.}~\bibnamefont {Laverock}}, \bibinfo {author}
  {\bibfnamefont {S.}~\bibnamefont {Dugdale}}, \bibinfo {author} {\bibfnamefont
  {M.}~\bibnamefont {Toney}},\ and\ \bibinfo {author} {\bibfnamefont
  {I.}~\bibnamefont {Fisher}},\ }\bibfield  {title} {\enquote {\bibinfo {title}
  {{Effect of chemical pressure on the charge density wave transition in
  rare-earth tritellurides RTe3}},}\ }\href@noop {} {\bibfield  {journal}
  {\bibinfo  {journal} {Physical Review B—Condensed Matter and Materials
  Physics}\ }\textbf {\bibinfo {volume} {77}},\ \bibinfo {pages} {035114}
  (\bibinfo {year} {2008})}\BibitemShut {NoStop}%
\bibitem [{\citenamefont {Forman}\ \emph {et~al.}(1972)\citenamefont {Forman},
  \citenamefont {Piermarini}, \citenamefont {Barnett},\ and\ \citenamefont
  {Block}}]{forman1972pressure}%
  \BibitemOpen
  \bibfield  {author} {\bibinfo {author} {\bibfnamefont {R.~A.}\ \bibnamefont
  {Forman}}, \bibinfo {author} {\bibfnamefont {G.~J.}\ \bibnamefont
  {Piermarini}}, \bibinfo {author} {\bibfnamefont {J.~D.}\ \bibnamefont
  {Barnett}},\ and\ \bibinfo {author} {\bibfnamefont {S.}~\bibnamefont
  {Block}},\ }\bibfield  {title} {\enquote {\bibinfo {title} {Pressure
  measurement made by the utilization of ruby sharp-line luminescence},}\
  }\href@noop {} {\bibfield  {journal} {\bibinfo  {journal} {Science}\ }\textbf
  {\bibinfo {volume} {176}},\ \bibinfo {pages} {284--285} (\bibinfo {year}
  {1972})}\BibitemShut {NoStop}%
\bibitem [{\citenamefont {Heid}\ and\ \citenamefont {Bohnen}(1999)}]{heid1999}%
  \BibitemOpen
  \bibfield  {author} {\bibinfo {author} {\bibfnamefont {R.}~\bibnamefont
  {Heid}}\ and\ \bibinfo {author} {\bibfnamefont {K.-P.}\ \bibnamefont
  {Bohnen}},\ }\bibfield  {title} {\enquote {\bibinfo {title} {Linear response
  in a density-functional mixed-basis approach},}\ }\href
  {https://doi.org/10.1103/PhysRevB.60.R3709} {\bibfield  {journal} {\bibinfo
  {journal} {Phys. Rev. B}\ }\textbf {\bibinfo {volume} {60}},\ \bibinfo
  {pages} {R3709(R)--R3712(R)} (\bibinfo {year} {1999})}\BibitemShut {NoStop}%
\end{thebibliography}
\end{document}